\documentclass[review,hidelinks]{elsarticle}
\usepackage{lineno}
\usepackage{xcolor}
\usepackage{hyperref}
\hypersetup{pdfauthor=author}
\usepackage{amsmath,amsfonts,amssymb}
\usepackage{caption}
\usepackage{subfigure}
\usepackage{tabularx,booktabs}
\usepackage{tabulary}
\usepackage{algorithm}
\usepackage{algpseudocode}
\usepackage{setspace}
\usepackage{pifont}
\usepackage{float}
\usepackage{booktabs}
\usepackage{multibib}
\usepackage{outlines}
\newcites{supp}{Supplementary References}
\newcommand{\cmark}{\ding{51}}%
\newcommand{\xmark}{\ding{55}}%

\modulolinenumbers[5]
\usepackage{geometry}[a4]

\journal{Imaging Neuroscience}

\biboptions{authoryear}

\begin{document}
\begin{frontmatter}

\title{Interpretable AI for relating brain structural and functional connectomes}

\author[mainaddress]{Haoming Yang\corref{correspondingauthor}}\cortext[correspondingauthor]{Corresponding author}
\ead{haoming.yang@duke.edu}

\author[mainaddress]{Steven Winter}
\ead{steven.winter@duke.edu}

\author[zhengwuaddress]{Zhengwu Zhang}
\ead{zhengwu_zhang@unc.edu}
\author[mainaddress]{David Dunson}
\ead{dunson@duke.edu}

\ead{}

\address[mainaddress]{Duke University, Durham, NC 27705, USA}
\address[zhengwuaddress]{University of North Carolina at Chapel Hill, Chapel Hill, NC 27599, USA}

\begin{abstract}One of the central problems in neuroscience is understanding how brain structure relates to function. Naively one can relate the direct connections of white matter fiber tracts between brain regions of interest (ROIs) to the increased co-activation in the same pair of ROIs, but the link between structural and functional connectomes (SCs and FCs) has proven to be much more complex. To learn a realistic generative model characterizing population variation in SCs, FCs, and the SC-FC coupling, we develop a graph auto-encoder that we refer to as Staf-GATE. We trained Staf-GATE with data from the Human Connectome Project (HCP) and show state-of-the-art performance in predicting FC and joint generation of SC and FC. In addition, as a crucial component of the proposed approach, we provide a masking-based algorithm to extract interpretable inferences about SC-FC coupling. Our interpretation methods identified important
SC subnetworks for FC coupling and relating SC and FC with sex.
\end{abstract}


\end{frontmatter}

\section{Introduction}
Central to the understanding of the human brain is studying the relationship between brain structure and functionality. 
In this paper, 
the structural connectome (SC) refers to the complete collection of white matter fiber tracts connecting different regions of the brain; the functional connectome (FC) refers to correlations in brain activity across regions of the brain, with activity measured with BOLD signals. Learning the link between the anatomy and functionality of the brain involves understanding the relationship between SC and FC, which is generally referred to as the SC-FC coupling problem \citep{Honey2009}. An early assumption to address this problem is that the directly connected regions of the brain are more correlated with their functional neural activation, which has been verified by numerous studies \citep{Koch2002, Skudlarski2008, Greicius2009, Honey2009, Damoiseaux2009}. However, much of the variation in functional connectivity cannot be explained by direct structural connections, suggesting that neurons are functionally connected through indirect structural connections \citep{Suarez2020}. 

Such indirect effects of SC connections have been studied using complex methods encompassing dynamic biophysical models, graph models, network communication models, and statistical learning models. With a set of differential equations linking SCs and neuron activity, biophysical models can generate synthetic time series of neuron activations, which can then be transformed into predicted functional connectomes \citep{Stephan2004, Deco2012, Wang2019}. Node distances, topological information, and graph harmonics of SC networks have been used to study functional connectivity \citep{Vertes2012, Preti2019}. Considering brain activation as information communicated between different regions of the brain through SC, network communication models consider communication efficiency as an essential factor in predicting FC \citep{Goni2014, Avena-Koenigsberger2018}. Statistical learning techniques including spatial autoregressive models (SAR) and network latent factor models have also been applied to the SC-FC coupling problem \citep{Messe2014, Misic2016}. However, these complex methods are still not flexible enough to accurately explain the connection between SC and FC: for example, the predicted FC generally can only explain 60\% of the variance in empirical FC.

A more flexible approach to the SC-FC coupling problem is deep learning. The multi-layer perceptron (MLP) of \cite{Sarwar2021} outperforms previous methods by achieving a group average correlation of 0.9 and an average individual correlation of 0.55. While it achieves state-of-the-art predictive performance, the MLP model lacks (1) a generative model to characterize the joint distribution of SC and FC, which empowers probabilistic inference of SC-FC coupling; and (2) the integration of important graph topology information of SCs, whose importance has been verified by numerous studies surveyed in \cite{Suarez2020}.

The generative aspect of deep learning models can be achieved through Variational Auto-Encoder (VAE)-based generative neural networks \citep{Kingma2014}, which have been applied to studies of connectomes. For example, \cite{Zhao2019} proposed a truncated Gaussian mixture VAE, which learns a lower dimensional representation of functional connectivity and identifies underlying clusters and outliers in FC. Integration of graph topology can be achieved through the Graph Auto-Encoder (GATE) and Regression GATE (reGATE) of \cite{Liu2021}. GATE/reGATE build upon network latent space generative models \citep{Hoff2002} using graph k-nearest neighbor layers to generate realistic SCs and model the joint probability distribution of SC and cognition traits. However, these VAE-based approaches focus on either SC or FC alone, and in general, the deep learning methods mentioned above lack interpretable inference due to the complexity of the model architectures. 

Motivated by the success of deep learning models for connectomes, we develop methodology for (1) including graphical features of SC to improve FC prediction, (2) characterizing the joint variation of SC and FC through a generative model, and (3) providing interpretable inference for SC-FC coupling. Leveraging the state-of-the-art predictive performance of MLP methods in SC-FC coupling and inspired by the flexibility of VAE-based methods, we developed the \textbf{St}ructural \textbf{a}nd \textbf{F}unctional \textbf{G}raph \textbf{A}u\textbf{t}o-\textbf{E}ncoder (Staf-GATE). To obtain interpretable inferences from 
Staf-GATE, we developed a perturbation-based algorithm to provide insights into the complex SC-FC coupling relationship. We summarize our main contributions as follows.

\begin{figure}%
\centering
\includegraphics[width=\textwidth]{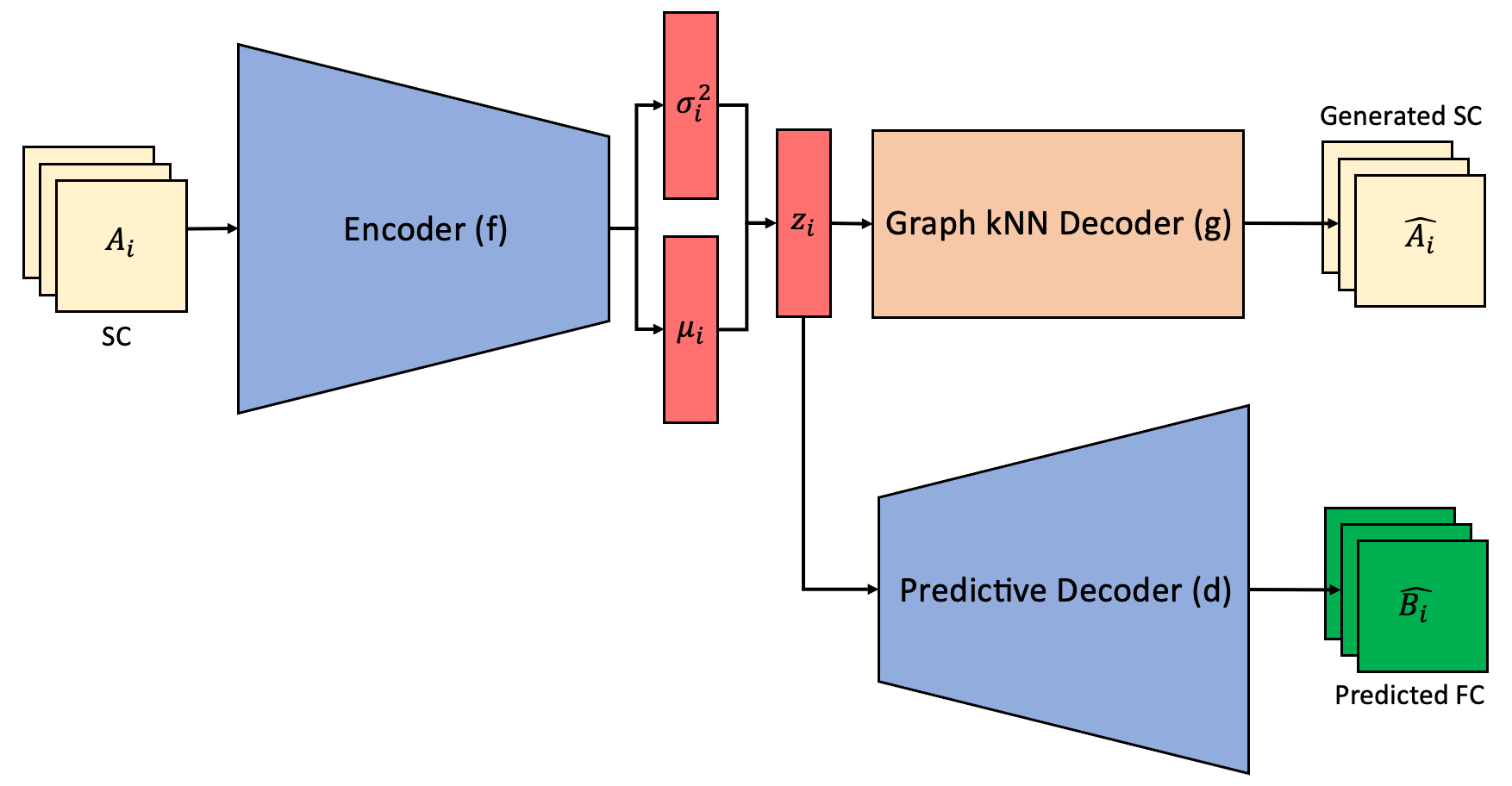}
\caption{The Staf-GATE Structure: The Encoder takes in input $A_i$, generates $z_i$, and then decodes $z_i$ to generate $\hat{A_i}$ again to learn the probability distribution of $A_i$. The predictive generator takes in latent parameter $\mu_i$ and predicts output $\hat{B_i}$}
\label{fig:Staf-GATE}%
\end{figure}

First, we learn the joint probability distribution of SC and FC through Staf-GATE, which consists of 3 components: an encoder, a decoder, and a predictive generator. Staf-GATE takes $A_i$, an individual's SC, as input and outputs denoised SC, denoted as $\hat{A_i}$ and predicted FC denoted as $\hat{B_i}$ (see Figure \ref{fig:Staf-GATE}). The encoder maps input $A_i$ to latent variables $z_i$. The decoder, which leverages graph k-nearest neighbor layers to incorporate the relative distance between nodes, learns the probability distribution of $A_i|z_i$ through a Poisson latent space model  \citep{Liu2021}. The predictive generator infers the latent variable $z_i$ targeted towards predicting output $B_i$. Collectively, the decoder and predictive generator learn the conditional joint distribution of $p(A_i, B_i|z_i)$.

Second, we provide interpretable inference through an iterative algorithm, which is inspired by the idea of interpreting neural networks using meaningful perturbations \citep{Fong2017}. We perturb the SC network by iteratively severing (masking) connections between pairs of ROIs and measuring the degradation in predictive performance. Edges critical for prediction are collected into a subnetwork that is crucial for SC-FC coupling. By repeatedly training Staf-GATE and running the interpretation algorithm, we found that our interpretation algorithm is robust and reproducible. As the algorithm is independent of Staf-GATE, it can be widely applied to other studies relying on deep learning to provide further inference.

We illustrate Staf-GATE's efficacy with data from the Human Connectome Project (HCP), showing significantly improved prediction of FC from SC over MLP models. Staf-GATE also generates  realistic SC and FC pairs that preserve the empirical network topology. The interpretation algorithm allows us to identify an important SC subnetwork for FC prediction, with fiber counts between the ROIs in the subnetwork strongly correlated with FC. We demonstrated the efficacy of our algorithm by exploring the difference in SC-FC coupling under different sex groups.
Our inference on male/female groups partially matches earlier findings that males have more within-hemisphere connections while females have more cross-hemisphere connections, but we also find that cross-hemisphere connections are equally important for males' SC-FC coupling. Code is available to reproduce our results and implement Staf-GATE at \href{https://github.com/imkeithyang/Staf-GATE.git}{github.com/imkeithyang/Staf-GATE.git}

\section{Materials and Methods}
\subsection{Data} \label{sec:data}
We use data derived from the HCP to illustrate and validate our methodology development. The HCP recorded high-resolution brain imaging data including diffusion MRI (dMRI) and functional MRI (fMRI) for more than 1000 outwardly healthy adults. HCP participants' basic information, such as sex and age, as well as behavioral traits, including oral reading ability and vocabulary ability, were also recorded. A detailed description of the data collection process can be found in \cite{VanEssen2013}. The data can be found on \href{http://www.humanconnectomeproject.org/data/}{humanconnectomeproject.org/}.

\subsubsection{Structural Connectome Mapping}
Each individual participating in the HCP was scanned by a customized 3T scanner to obtain dMRI data. Individuals were scanned from left-to-right and right-to-left encoding directions with the following scanning parameters: multiband factor of 3, 1.25$\text{mm}^3$ voxel size, a total of 270 diffusion weighting directions equally distributed across b-values of 1000, 2000, 3000 s/$\text{mm}^2$ \citep{VanEssen2013}. The dMRI data were then pre-processed through the minimal pre-processing pipeline developed in \cite{Glasser2013} including correcting the eddy current induced field inhomogeneities, head motions, and gradient-nonlinearity distortion. The corrected data were then transformed from native space to structural volume space with gradient vectors rotated to align with the direction in structural space.

Our data were kindly provided by \cite{Sarwar2021}, and they applied the following steps for whole-brain tractography and structural connectome preprocessing. They estimated the fiber orientation in each white matter voxel using constrained spherical deconvolution (CSD) with a set of 8-order spherical harmonics  \citep{Tournier2007}.
A white matter mask, derived from automated structural segmentation, was applied to generate streamlines. A one-voxel dilation of the mask boundaries was applied to the white matter mask to cover gaps between grey and white matter boundaries. \cite{Sarwar2021} then used the sd\textunderscore{stream} option in the tckgen function of the MRtrix package with default parameters of step size ($0.1 \times \text{voxel size}$), angle threshold ($9^{\circ} \times \text{step size}/\text{voxel size}$), and FOD threshold (0.1) to propagate the streamlines through the estimated orientation, generating 2 million streamlines for each subject, with a maximum streamline length of 400 mm. Finally, the number of streamlines connecting each pair of regions under the Desikan-Killiany parcellation \citep{Desikan2006} was mapped to structural connectivity matrix of each subject. We refer the readers to \cite{Sarwar2021} for the detailed documentation of their preprocessing.

Before training the neural network, \cite{Sarwar2021} applied a Gaussian resampling preprocessing developed by \citep{Honey2009} on SCs to reduce the range of SC elements for a more stable neural network training. We found that such preprocessing, as well as the more common log-transformation, will distort the SCs' graph topological features. To ensure training stability but preserve the graph's topological features, we chose to scale the elements of the SCs down by 100. Further discussion of different preprocessing methods is deferred to Supplement \ref{appendix:preprocess}.

\subsubsection{Functional Connectome Mapping}
The HCP participants were also scanned for their resting state fMRI. fMRI data were collected through a 15-minute scan for each encoding direction (left-to-right and right-to-left) with the following scanning parameters: multiband factor of 8, 2 $\text{mm}^3$ voxel size, and a TR of 0.7s. The resting state fMRI data also underwent a minimal pre-processing pipeline developed by \cite{Glasser2013}. The pre-processing steps include removing spatial distortion, correcting head motion, and normalization. 

The functional connectomes are represented as Pearson correlation matrices of BOLD signals between regions. The minimal pre-processed fMRI of an individual is comprised of voxel-specific BOLD activation time series. The voxel-specific BOLD activations within an ROI were averaged to construct the regional activation time series. Then the correlation of activation between regions $u,v$ is computed to form element $(u,v)$ of the FC matrix. 

\subsection{The Structural and Functional Graph Auto-Encoder}
We propose the \textbf{St}ructural \textbf{a}nd \textbf{F}unctional \textbf{G}raph \textbf{A}u\textbf{t}o-\textbf{E}ncoder (Staf-GATE). 
Staf-GATE consists of 3 main parts: encoder, decoder, and predictive generator. The encoder maps the structural connectome (SC) input, $A_i$, to latent parameters, $\mu_i$ and $\sigma_i^2$, for subject $i$. A random sample of latent variable $z_i$ from $N(\mu_i, \sigma_i^2)$ is fed into the decoder to generate a sample of $\hat{A_i}$. The predictive generator uses the same latent variable sample $z_i$ to generate the functional connectome (FC) output, $\hat{B_i}$, trying to predict the empirical FC $B_i$. The structure of Staf-GATE is presented in Figure \ref{fig:Staf-GATE}. We denote the encoder as $f$, the decoder as $g$, and the predictive generator as $d$. 

We will next develop generative models for the decoders(generators) to model the sparse SC and dense FC flexibly. Once the generative models for SC and FC are defined, our next step involves calculating the Evidence Lower Bound (ELBO) to approximate the joint likelihood of SC and FC. This approximation is crucial for training Staf-GATE through variational inference. It's important to highlight that the methodologies employed in Staf-GATE are influenced by a variety of related studies. For a comprehensive discussion on these connections, please refer to Supplement \ref{appendix:connections}, which has been provided to keep the main paper succinct.

\subsubsection{Staf-GATE generative models}
\label{sec:gen_model_detail}
We develop the generative model for decoder output $g(z_i)$ and predictive generator output $d(z_i)$. We use the notation $M_i \in \mathbf{R}^{V \times V}$ to represent an arbitrary $V$ node undirected weighted graph's adjacency matrix, $i\in {1,2,...,N}$. The edge between node $u,v$ is denoted as $M_{i[uv]}$.  Adjacency and correlation matrices are both symmetric; therefore, it suffices to model the lower triangular elements as
    $L(M_i) = (M_{i1},\ldots, M_{iV(V-1)/2}) 
    \equiv (M_{i[21]},M_{i[31]},\ldots, M_{i[V1]}, M_{i[V2]}, ...M_{i[V(V-1)]})$.

The Staf-GATE encoder $f$ maps SC to $K$ dimensional latent variable $z_i \in \mathbf{R}^{K}$ through a neural network and then passes $z_i$ to the decoder $g$ to generate realistic SCs. We assume that the fiber counts in SC between ROI pairs (i.e., elements in $A_i$) are conditionally independent Poisson random variables given Poisson rates $\Lambda_i$ \citep{Hoff2002}. The process can be mathematically written as: 
\begin{gather}
    f:\mathbf{R}^{V(V-1)/2} \to \mathbf{R}^{K};\quad L(A_i)\mapsto z_i \\
    g:\mathbf{R}^{K} \to \mathbf{R}^{V \times V};\quad z_i\mapsto \hat{A_i} \\
    \hat{A_i} \sim \text{Poisson}(\Lambda_i); \quad \Lambda_i \in \mathbf{R}^{V \times V}
\end{gather}

We further decompose $\log(L(\Lambda_i))$ into a global edge-specific rate $\gamma \in \mathbf{R}^{V(V-1)/2}$ and a subject-specific deviation $\psi^{(i)}(z_i) \in \mathbf{R}^{V(V-1)/2}$ as shown in Equation \eqref{eq:lambd_decomp}. The global edge-specific Poisson rates are shared across the population; the subject-specific deviations are modeled as a function of the latent variable $z_i$. For $A_i \in \mathbf{R}^{V \times V}, V=68$, the resulting lower triangular component $L(A_i) \in \mathbf{R}^{V(V-1)/2=2278}$ remains high dimensional. For dimension reduction in modeling $A_i|z_i$, we adapt the 
 latent factor model of \cite{Durante2017} as in Equation \eqref{eq:latent_factor}--\eqref{eq:one_factor}, with $r = 1,\ldots,R$ indexing the latent factor. To summarize: 
\begin{gather}
    \log(\Lambda_{i}) = \gamma + \psi^{(i)}(z_i) \label{eq:lambd_decomp} \\
    \psi^{(i)}(z_i) = \sum_{r=1}^R \alpha_r X_r^{(i)}(z_i)X_r^{(i)}(z_i)^\intercal
    \label{eq:latent_factor} \\ 
    X^{(i)}_r(z_i) = (X^{(i)}_{1r}(z_i),..., X^{(i)}_{Vr}(z_i))^\intercal \in \mathbf{R}^{V}
    \label{eq:one_factor}
\end{gather}

To generate realistic SCs, which typically exhibit sparsity and specific network topologies, we incorporate graph k-nearest neighbor (kNN) layers into our decoder. These layers are utilized to model the latent factors, denoted as $X^{(i)}_r$, as suggested by \cite{Liu2021}. The distance $D_{[u,v]}$ between two regions $u,v$ is defined to be inversely related to the number of fibers connecting (the higher number of connections the lower the distance) these regions with (as a convention) $D_{[u,u]} = 0$ and $D_{[u,v]} = \infty$ when $u$ and $v$ are unconnected. With this notion of distance, the $k$-nearest neighbors of a region $u$ are the $k$ regions that have the largest fiber count among all of $u$'s connected regions.

An M-layer Graph-kNN neural network can be formally defined as:
\begin{gather}
    X^{(i,1)}_r = h_1(W_1 z_i + b_1), \; \text{for } m=1 \\
    X^{(i,m)}_r = h_m(W_m X^{(i,m-1)}_r(z_i) + b_m), \; \text{for } 2 \le m \le M, 
\end{gather}
where $X^{(i,m)}_r$ is the output of the $m$th graph-kNN layer, and $h_m$ is the activation function of the $m$th layer. The set of weights of the $m$th layer is denoted as $W_m$, and is masked by the kNN, meaning that only the weights of $k$-nearest neighbors of a node are non-zero. The kNN mask preserves the top k strongest connections between ROIs and provides the sparsity needed for generating SC. The kNN mask affects FC construction indirectly since it impacts the mapping from $A_i$ to $z_i$.  For including the complete topology of the graph, we model latent factor $X^{(i)}_r$ with a $k=2^r$ mask to encode different levels of connections in learning the latent factors.

\begin{figure}
    \centering
    \includegraphics[width=\textwidth]{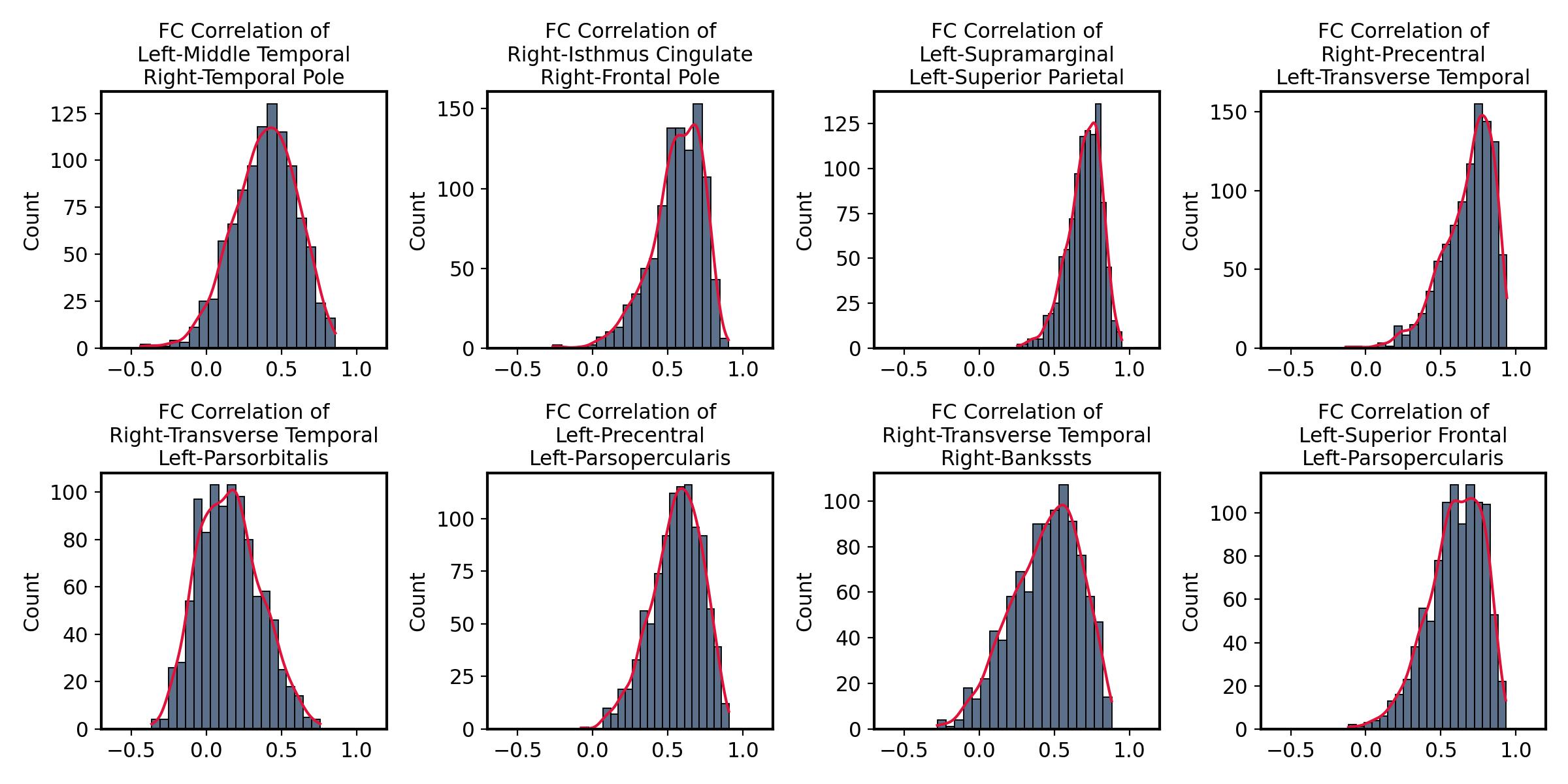}
    \caption{Histograms of FC correlations of selected ROI pairs. The histograms are generally skewed; therefore, using skew-Normal distribution can better characterize FCs' population variation.
    }
    \label{fig:skew_FC}
\end{figure}

It remains to specify a predictive generator for the FC $B_i$.
Similar to previous applications of VAEs for FC data, 
 such as \cite{Zhao2019} and \cite{Kim2021}, we specify a generative model for $B_i$ given latent variable $z_i$; we infer $z_i$ only from the SC $A_i$.
 Although \cite{Zhao2019} assumed a Gaussian distribution for the elements of $B_i$, Figure \ref{fig:skew_FC} shows that the elements have a skewed distribution, motivating a skewed Gaussian model \citep{Azzalini1985}: 
\begin{gather}
    x \sim \text{skew-Gaussian}(\xi, \omega, \kappa) \nonumber\\
    p(x) = 2 \phi \left(\frac{x-\xi}{\omega} \right)\Phi\left[\kappa \left(\frac{x-\xi}{\omega}\right)\right] \label{eq:skew}\\
    \text{where } \phi, \Phi \text{ are the PDF and CDF of } N(0,1) \nonumber
\end{gather}
Due to symmetry of $B_i$, we focus on the lower-triangular vector 
 $L(B_i) \in \mathbf{R}^{V(V-1)/2}$. We assume that the entries in $L(B_i)$ are conditionally independent given latent variable $z_i$, with $L(B_i)|z_i \sim \text{skew-Gaussian}(d(z_i), \text{diag} \left\{\Omega_B^2 \right\}, \kappa_B)$, $d(z_i)$ the predictive generator output, $\Omega_B^2 \in \mathbf{R}^{V(V-1)/2}$ scale parameters specific to each element of $L(B_i)$, 
and $\kappa_B \in \mathbf{R}^{V(V-1)/2}$ a skewness parameter. We estimate $\Omega_B^2$ applying the method of moments estimator of \cite{Azzalini1999} to the training data, and optimize $\kappa_B$ as a neural network parameter.

\subsubsection{Evidence Lower Bound and Variational Inference}
With the generative models developed in the previous section, we collect the parameters into a vector $(\gamma, \alpha_r, W_m, b_m, \kappa_B)$ for $r = 1, \ldots,R,\; m = 1,\ldots,M$ as $\theta$. The training of Staf-GATE aims to minimize the negative log likelihood of joint distribution $p_{\theta}(A_i, B_i)$ and estimate the posterior distribution of the latent variables $p_{\theta}(z_i|A_i, B_i)$.

The distributions $p_{\theta}(A_i, B_i)$ and $p_{\theta}(z_i|A_i,B_i)$ are intractable; therefore, we use variational inference (VI) \citep{Jordan1999} to train our neural network. In particular, we approximate the latent distribution $p_{\theta}(z_i|A_i, B_i)$ with $q_{\phi}(z_i|A_i) = N(\mu_{\phi}(A_i),\sigma_{\phi}(A_i) I_K)$, with $\mu_{\phi}(A_i), \sigma_{\phi}(A_i)$ parameters output by encoder $f(A_i)$. We minimize the difference between $q_{\phi}(z_i|A_i)$ and $p_{\theta}(z_i|A_i, B_i)$ by minimizing the KL-divergence between the two distributions defined in Equation \eqref{eq:kld}. We show in Equation \eqref{eq:kld}-\eqref{eq:loglike} that  minimizing $D_{KL}(q_{\phi}(z_i|A_i) || p_{\theta}(z_i|A_i, B_i))$ and $-\log p_{\theta}(A_i, B_i)$ can be achieved by minimizing the negative evidence lower bound (ELBO) denoted as $\mathcal{L}$.
\begin{align}
D_{KL}(q_{\phi}(z_i|A_i) || p_{\theta}(z_i|A_i, B_i)) &= \mathbf{E}_{Z \sim Q}\left[\log\frac{q_{\phi}(z_i|A_i)}{p_{\theta}(z_i|A_i, B_i)}\right] \label{eq:kld} \\
&= \mathbf{E}_{Z \sim Q}\left[\log \frac{q_{\phi}(z_i|A_i)p_{\theta}(A_i, B_i)}{p_{\theta}(z_i, A_i,B_i)}\right] \nonumber \\
&= \mathbf{E}_{Z \sim Q}\left[\log p_{\theta}(A_i, B_i) + \log \frac{q_{\phi}(z_i|A_i)}{p_{\theta}(z_i, A_i,B_i)}\right] \nonumber \\
&= \log p_{\theta}(A_i, B_i) \underbrace{- \mathbf{E}_{Z \sim Q} \left[ \log \frac{p_{\theta}(z_i, A_i,B_i)}{q_{\phi}(z_i|A_i)} \right]}_{\mathcal{L}(A_i, B_i, \theta, \phi)} \nonumber \\
\mathcal{L}(A_i, B_i, \theta, \phi) &= D_{KL}(q_{\phi}(z_i|A_i) || p_{\theta}(z_i|A_i, B_i))-\log(p_{\theta}(A_i, B_i)) \label{eq:loglike}\\
\mathcal{L}(A_i, B_i, \theta, \phi) &= -\mathbf{E}_{Z \sim Q} \left[ \log \frac{p_{\theta}(z_i, A_i,B_i)}{q_{\phi}(z_i|A_i)} \right] \nonumber \\ &= -\mathbf{E}_{q_{\phi}(z_i|A_i)}\log p_{\theta} (B_i | z_i) - \mathbf{E}_{q_{\phi}(z_i|A_i)}\log p_{\theta} (A_i | z_i) \nonumber \\
& + D_{KL}(q_{\phi}(z_i|A_i)||p_{\theta}(z_i)) \label{eq:ELBO}
\end{align}

The ELBO can be decomposed into three terms as presented in Equation \eqref{eq:ELBO}. The first term, $\mathbf{E}_{q_{\phi}(z_i|A_i)}\log p_{\theta} (B_i | z_i)$, serves as the supervised reconstruction error for FC generation, which decreases as the generated FCs better characterizes the empirical FC distribution. The second term, $\mathbf{E}_{q_{\phi}(z_i|A_i)}\log p_{\theta} (A_i | z_i)$, is the self-supervised reconstruction error that measures how well the model can recover the empirical SC distribution. The third term is the KL divergence between the latent variable $z_i$'s approximated posterior and its prior, which serves as a regularization to shrink the approximated posterior $q_{\phi}(z_i|A_i) = N(\mu_i(A_i), \sigma_i^2(A_i)I_K)$ towards the prior $p(z_i) = N(0, I_k)$. We minimize $\mathcal{L}$ to train the neural network for approximating the posterior of $z_i|A_i$ and the joint distribution of $A_i, B_i$. In practice, the expected values in $\mathcal{L}$ are intractable, and they are approximated by a Monte Carlo approximation through repeated sampling of $z_i \sim N(\mu_i(A_i), \sigma_i(A_i))$.

\subsubsection{Regularization Formulation}
In this section, we augment the loss function with an additional regularization term related to the one in \cite{Sarwar2021} to improve performance in realistically characterizing variability across individuals. We denote empirical SC as eSC, generated SC as gSC, empirical FC as eFC, and generated FC as gFC.
Regularization aims to ensure Staf-GATE does not simply predict mean FC and forces the gFCs to retain the same variation as eFCs. We measure variation of FCs by summing the Pearson correlation between FCs (inter-FC correlation): 
\begin{gather}
    \text{inter-FC correlation} = \rho_{\text{FC}} = \sum_i^n\sum_{j \neq i}^n \text{r}(\text{FC}_i, \text{FC}_j) \label{eq:corr_reg}
\end{gather}
Our regularization term is $\lambda (\rho_{\text{gFC}}-c)^2$, where $\rho_{\text{gFC}}$ is the inter-gFC correlation. Hyperparameters $c$ and $\lambda$ will be tuned through a grid search. Combining the ELBO from our previous derivation and our regularization, the complete loss is: 
\begin{gather}
\mathbf{L} = \mathcal{L} + \lambda (\rho_{\text{gFC}} - c)^2 \label{eq:loss}
\end{gather}

\section{Simulation Study}
\label{sec:simulation}
In this section, we test performance on simulated data containing multiple groups of topologically distinct networks. Simulated SCs are denoted as $\tilde{A}_{yi}$ where $y=1,2,3,4$ indexes the group and $i=1,...,N_y$ indexes the $N_y$ individuals in that group. All simulated SCs have the same number of edges, $|\tilde{A}_{yi}|$=1350.\footnote{The average number of SC connections excluding the diagonal elements in the adjacency matrices is 1348.} We decomposed each simulated adjacency matrix as the sum of the group level edges and an individual level perturbation: $\tilde{A}_{yi} = \tilde{A}_y + E_{yi}$. 

In order to make the simulations as realistic as possible while maintaining distinguishable group differences, the group edges $\tilde{A_y}$ are chosen by randomly selecting $|\tilde{A}_y|$ edges from the most common SC edges in the HCP data. We take $|\tilde{A}_y|=y\times 50$, so the first group-level subnetwork has 50 connections and the fourth has 200. The individual perturbations $E_{yi}$ consist of $|\tilde{A}_{yi}| - |\tilde{A}_y|$ edges chosen at random from the set of pooled SC edges, which is the set of all possible edges excluding the edges in $\tilde{A}_y$. 
Figure \ref{fig:simu_SC_example} shows examples of simulated networks, as well as violin plots of topological summaries by groups. The network density, clustering coefficient, and eigen centrality increase as the number of group edges increases. 

We simulate a corresponding population of FCs for each group. We start by simulating BOLD time series of length $T=100$ across brain regions
for each individual; the elements of the FC matrices are correlation coefficients in BOLD series between pairs of brain regions.
Letting $\Upsilon_{yi}\in R^{V\times T}$ denote the BOLD time series for individual $i$ in group $y$, we let
\begin{equation}
    \Upsilon_{yi} = Q_y^{-1}(\tilde{A}_{yi} \eta_i + \epsilon_i), \label{eq:simu_bold}
\end{equation}
where $\tilde{A}_{yi}$ is the simulated SC for individual $i$, $\eta_i\sim N(0, I)$ is a latent factor, $\epsilon_i\sim N(0, I)$ is noise, and $Q_y$ is a group perturbation matrix with constraint $||Q_y-I||<0.8$ described in \cite{Roy2021}; this induces a different group topological structure on the simulated FC from the simulated SC. We denote the resulting simulated FC as $\tilde{B}_{yi}$ which is the correlation between different rows in $\Upsilon_{yi}$; group averages are presented in Figure \ref{fig:simu_FC_example} together with topological summaries. Each group has a different topology structure.

\begin{figure}
    \centering
    \subfigure[Predicted group average FC vs true group average FC]{\includegraphics[width=\textwidth]{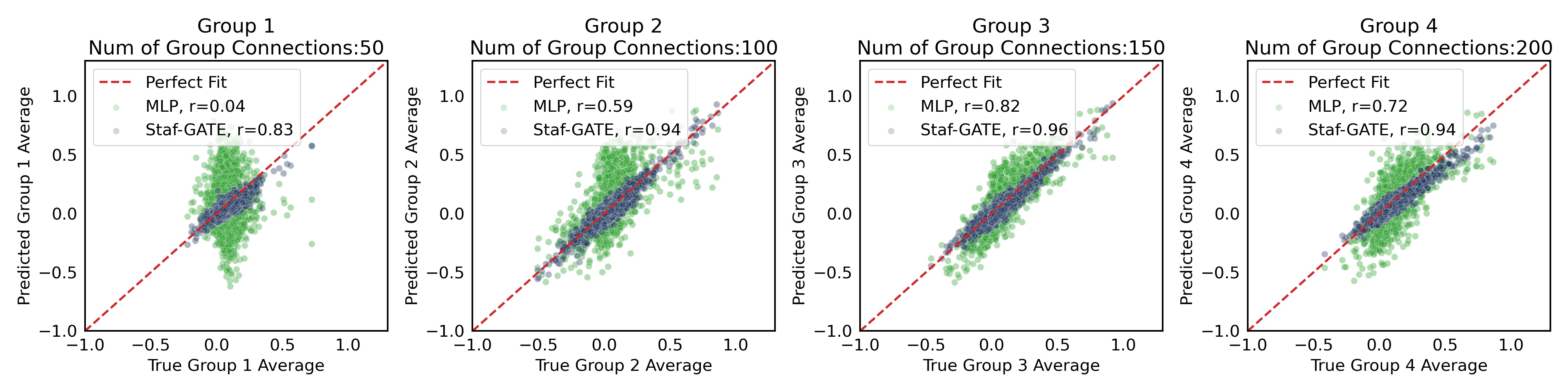}
    \label{fig:simu_scatter}}
    \subfigure[Comparison of selected network statistics of predicted $\tilde{B}_{yi}$ and true $\tilde{B}_{yi}$]{\includegraphics[height=\textwidth]{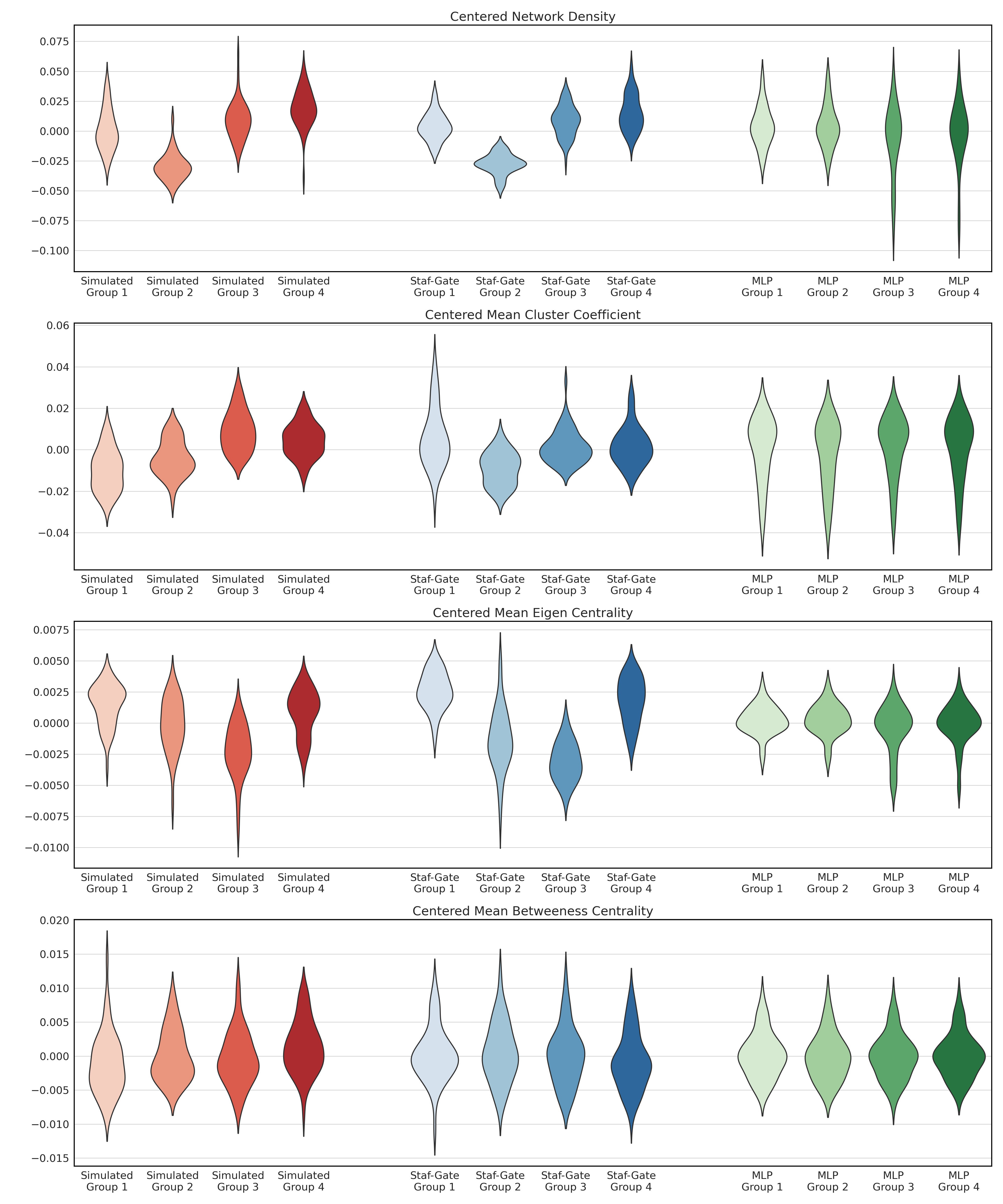}
    \label{fig:simu_pred_violin}}
    \caption{Results of simulation study. \textbf{a)} Predicted group average FC vs true group average FC for different groups at different levels of group signals; the dotted red line indicates the line of perfect fit. \textbf{b)} Comparison of selected network statistics of predicted $\tilde{B}_{yi}$ and true $\tilde{B}_{yi}$. Each set of violin plots corresponds to a network statistic. The colors red, blue, and green correspond to the true simulated data, Staf-GATE predictions, and MLP predictions; to better illustrate the differences in topology between groups, each distribution of the summary statistic is centered by subtracting the population mean of that statistic.}
\end{figure}

We trained Staf-GATE and compared it to baseline MLP \citep{Sarwar2021} with 800 generated $\tilde{A}_{yi}, \tilde{B}_{yi}$ pairs and evaluated the predictive ability of the different approaches with 200 generated pairs as the test set. We trained both Staf-GATE and the MLP model in \citep{Sarwar2021} for 1000 epochs with the Adam optimizer using a learning rate of $1e-4$. The training batch size for Staf-GATE and MLP is 200 and 5, correspondingly. We evaluated the goodness of fit with group average correlation and mean squared error. Figure \ref{fig:simu_scatter} shows 
the predicted group average of the $\tilde{B}_{yi}s$ vs. the true group average. Staf-GATE outperforms MLP for all four groups in terms of group average correlation. MLP exhibits a clear lack of fit and struggles with high noise inputs, in contrast to Staf-GATE.

We are interested in studying Staf-GATE's ability to leverage the group topological differences in $\tilde{A}_{yi}$ to predict better the network topology in $\tilde{B}_{yi}$. Figure \ref{fig:simu_pred_violin} presents network topology statistics of FCs. MLP predictions fail to discern the group differences in network topology. Staf-GATE, however, can recover the topological differences between groups. Moreover, Staf-GATE effectively learns low-dimensional brain network representations, as demonstrated in Figure \ref{fig:latent_rep}. This enables the identification of group structure, clusters, and outliers in the brain network data, which may not be evident from direct examination of the adjacency matrices of SCs and FCs. More details are in Supplement \ref{appendix:simu}.

\section{Results for HCP Data}
\label{sec:results}
We compared Staf-GATE to the state-of-the-art (SOTA) MLP model in \cite{Sarwar2021} on data from the HCP. Despite the original MLP's good results being achieved with resampled SC - a method which alters SC's network topology and hinders our aim of examining SC-FC coupling - we maintained fairness by comparing Staf-GATE to two MLP versions: one trained with resampled SC (MLP resample), and the other with scaled SCs (MLP scaled).

We denote an empirically observed FC as eFC and a predicted FC as pFC. We split the 1000 SC-FC pairs into a 900-100 training-test set. Pairs of twins were split into the same dataset. The three hyperparameters -- the dimension of $z_i$ denoted as $K$, regularization parameters $\lambda$, and $c$ -- were tuned alongside the learning rate and batch size through grid search. Staf-GATE is trained with the best parameter settings from the grid search for 5000 epochs. The Staf-GATE predictive generator's activation function was chosen to be Tanh because the correlation between nodes can be negative. Full details regarding tuning are in Supplementary \ref{appendix:tune}. The complete structure of the neural network as well as detailed model comparisons including training complexity and functionality can be found in Tables \ref{tab:nn} and \ref{tab:comp}. 

\begin{table}
\centering
\caption{Staf-GATE architecture for different components. I=O=2278 is the input and output size; K=68 is the latent variable dimension; M=2 is the number of Graph kNN layers in the decoder; R=5 is the number of latent factors in the SC generative model. The last row of the table is the activation function for that component, and the last layer of a particular component only has identity activation.} 
\label{tab:nn} 
\footnotesize
\begin{tabularx}{\textwidth}{l*{6}{X}}
 \multicolumn{2}{l}{Encoder} & \multicolumn{2}{l}{Decoder} & \multicolumn{2}{l}{Predict-Generator} \\
\toprule
$\mu$  & $\sigma$ & Layer & Weight & Layer & Weight \\ 
\midrule
I $\times$ 1024 & I $\times$ 1024  &  Linear      & K $\times$ 68  & Linear & K $\times$ 128 \\ 
1024$\times$128 & 1024$\times$128  &  Graph kNN   & 68 $\times$ 68 & Linear & 128$\times$ 512 \\ 
128 $\times$ K  & 128 $\times$ K   &  k=$2^r$     & M = 2          & Linear & 512$\times$1024 \\
                &                  & $r\in 1,\ldots,R$ & R = 5          & Linear & 1024$\times$ O\\ 
Bias=True       &  Bias=True       &              &  Bias=True     & & Bias=False \\ 
Relu            & Relu             &              &  Sigmoid       & & Tanh \\
\bottomrule
\end{tabularx}
\end{table}

\begin{table}
\footnotesize
\centering
\caption{Comparison of performance and functionality: MLP vs Staf-GATE} \label{tab:comp}
\begin{tabularx}{\textwidth}{l*{4}{X}}
\toprule
& MLP Resample  & MLP Scale & Staf-GATE \\
\midrule
Group Avg Corr  & 0.90 & 0.71 & \textbf{0.96} \\
Avg Individual Corr  &  0.548 & \textbf{0.571} & \textbf{0.572} \\
Train Time & 19 Hours & 19 Hours & \textbf{25 minutes}\\
Train Epochs & 20,000 & 20,000 & \textbf{0}\\
Data Generation & \xmark & \xmark & \cmark \\
Low-Dim Representation & \xmark & \xmark & \cmark \\
\bottomrule
\end{tabularx}
\end{table}

\begin{figure}
\centering
\includegraphics[width=\textwidth]{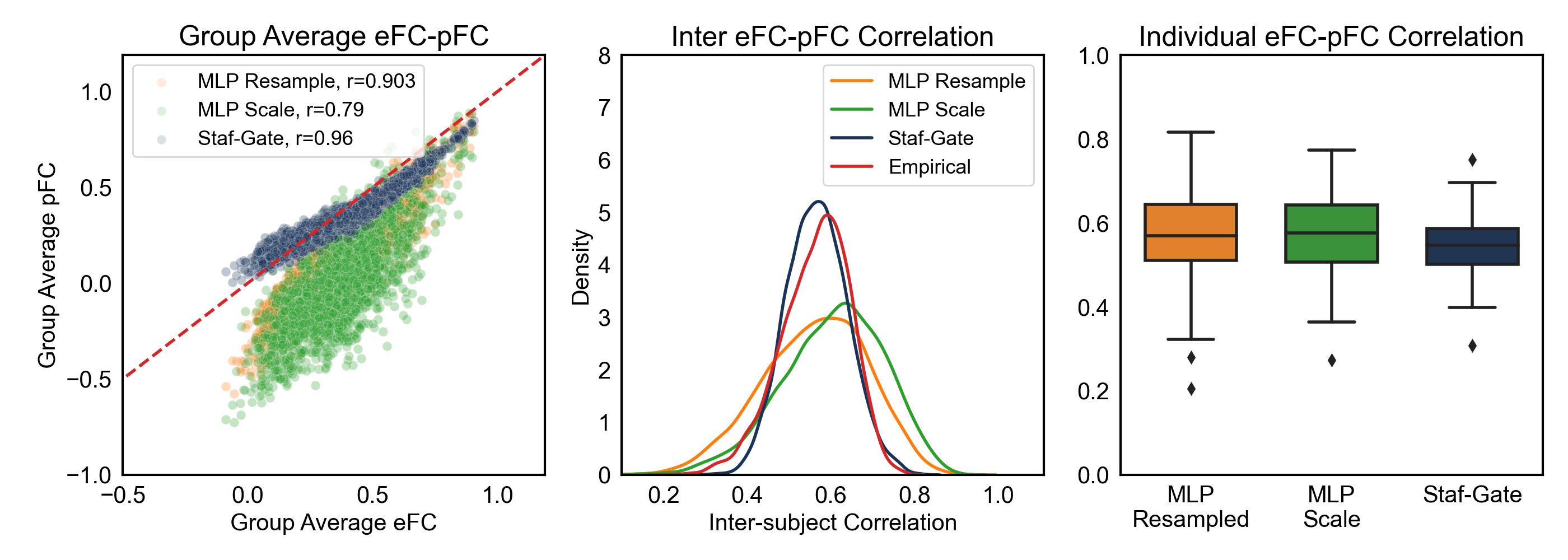}
\caption{Goodness-of-fit comparison for FC prediction between MLP and Staf-GATE. Left: Group average predicted functional connectome versus empirical functional connectome values; the red dashed line indicates perfect fit. Middle: The inter-subject correlation between all possible pairs within the predicted FC and empirical FC in the test set. Right: Boxplots comparing the correlation between empirically observed FC and predicted FC at an individual level (MLP Scale mean correlation=0.571, Staf-GATE mean correlation = 0.572).}
\label{fig:FC_corr}
\end{figure}

We assessed group-level goodness-of-fit through correlations and network summary statistics. Traditionally, SC-FC coupling results were evaluated via group average pFC-eFC correlation \citep{Sarwar2021}. Staf-GATE obtained a correlation of 0.96, a 6.6\% improvement over MLP resample, and 35\% improvement over MLP. As Figure \ref{fig:FC_corr} left panel presents, Staf-GATE's predicted group average follows the line of best fit nearly perfectly, whereas MLP's predictions are mostly below the line of best fit. Figure \ref{fig:FC_corr} middle panel compares the distribution of inter-pFC correlation produced by different methods against the inter-eFC correlation (which is denoted as Empirical and plotted in red color). We can see that Staf-GATE captures inter-subject variation accurately compared with the empirical distribution, whereas MLP can overestimate the inter-eFC correlation. Individual specific eFC-pFC correlations are presented in Figure \ref{fig:FC_corr} (right). Compared with the MLP methods, Staf-GATE's result has a smaller variance but a slightly lower mean. 

\begin{figure}
    \centering
    \includegraphics[width=1\textwidth]{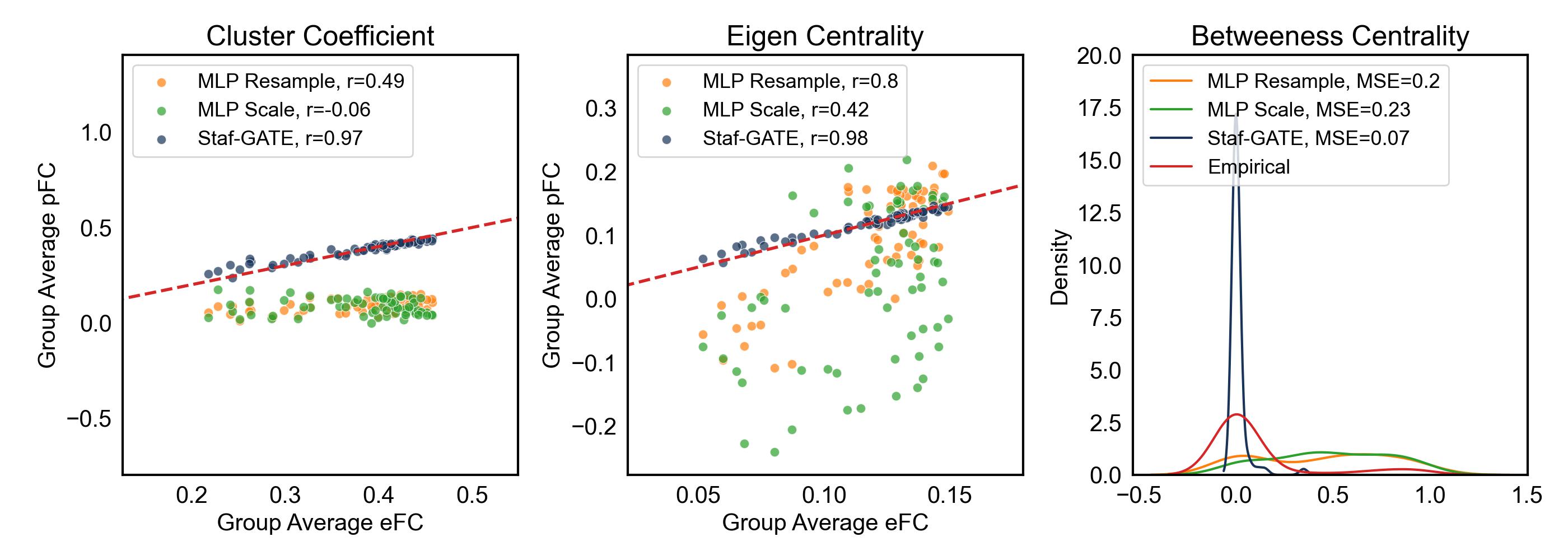}
    \caption{Network topology goodness-of-fit comparison between FC group averages predicted by MLP, Staf-GATE, and the empirical FC group average. The red dashed line in the left and middle panels indicates a perfect fit. The empirical node-level betweenness centrality is sparse, hence a correlation between the empirical and predicted is unreliable. We instead plot the density of node-level betweenness centrality and calculate the MSE in the right panel.}
    \label{fig:group_avg_topology}
\end{figure}

However, the correlation between eFC and pFC can only partially evaluate the performance of the proposed model. Maintaining the topology structures in predicted networks is also critical for network data prediction. Consequently, we also compared network summary statistics, including the clustering of nodes and different centrality measures.
Figure \ref{fig:group_avg_topology} compares the node-level network summary statistics between group averages of pFCs and eFCs, showing that Staf-GATE's prediction accurately recovers the network topology, but MLP models do not.

\begin{figure}
\centering
\subfigure[Goodness-of-fit assessment of generated SC.]{
\includegraphics[width=0.75\textwidth]{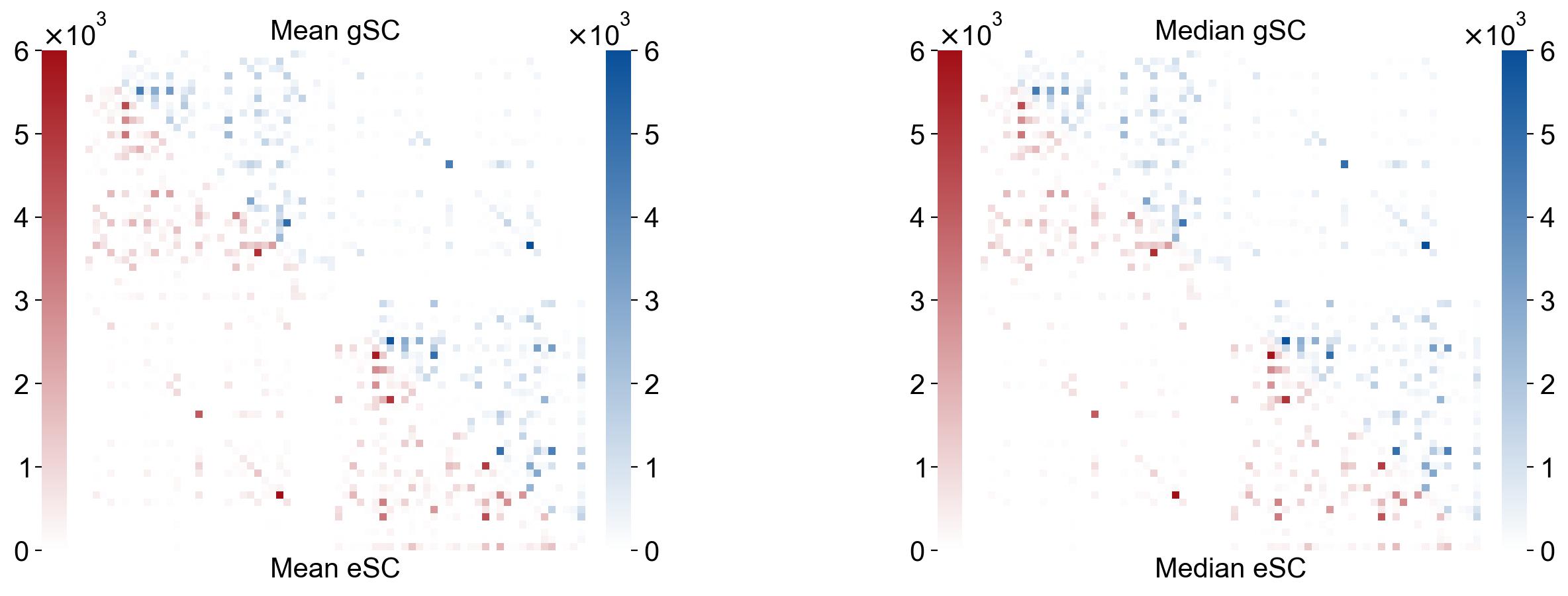}
\label{fig:gen_SC}}
\subfigure[Correlation goodness of fit of group average generated FC.]{
\includegraphics[width=0.75\textwidth]{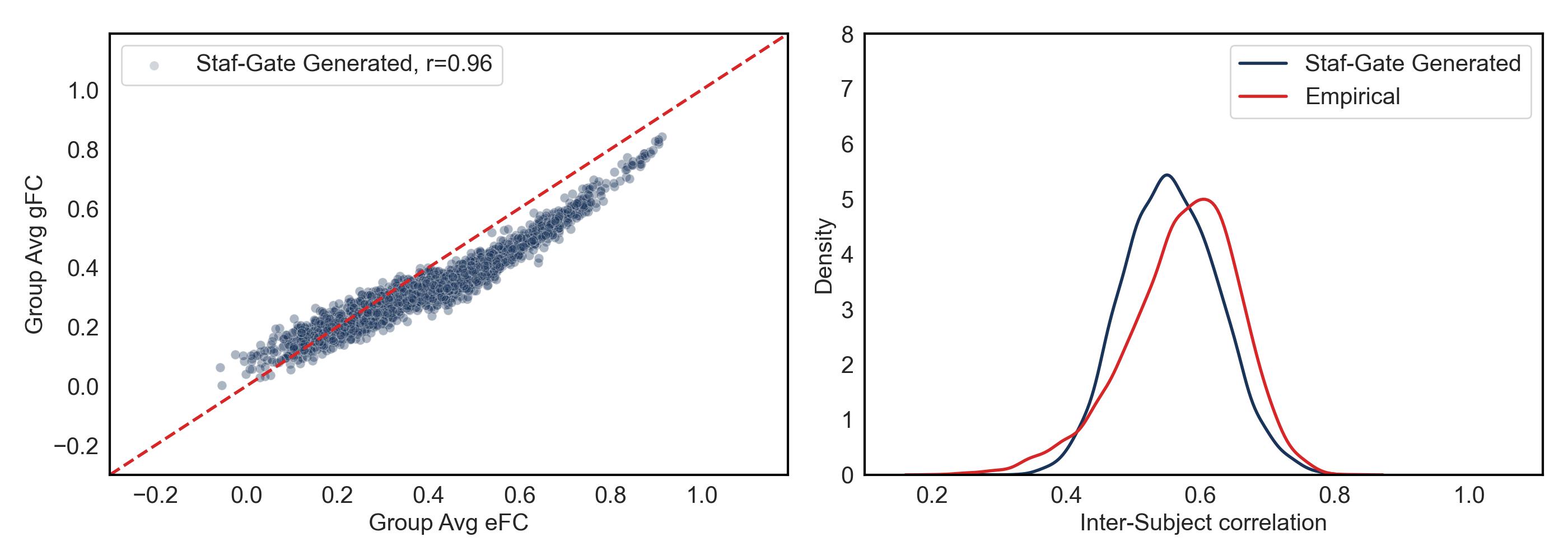}
\label{fig:fc_gen_corr}}
\subfigure[Network topology goodness-of-fit assessment for generated SC and FC.]{
\includegraphics[width=0.75\textwidth]{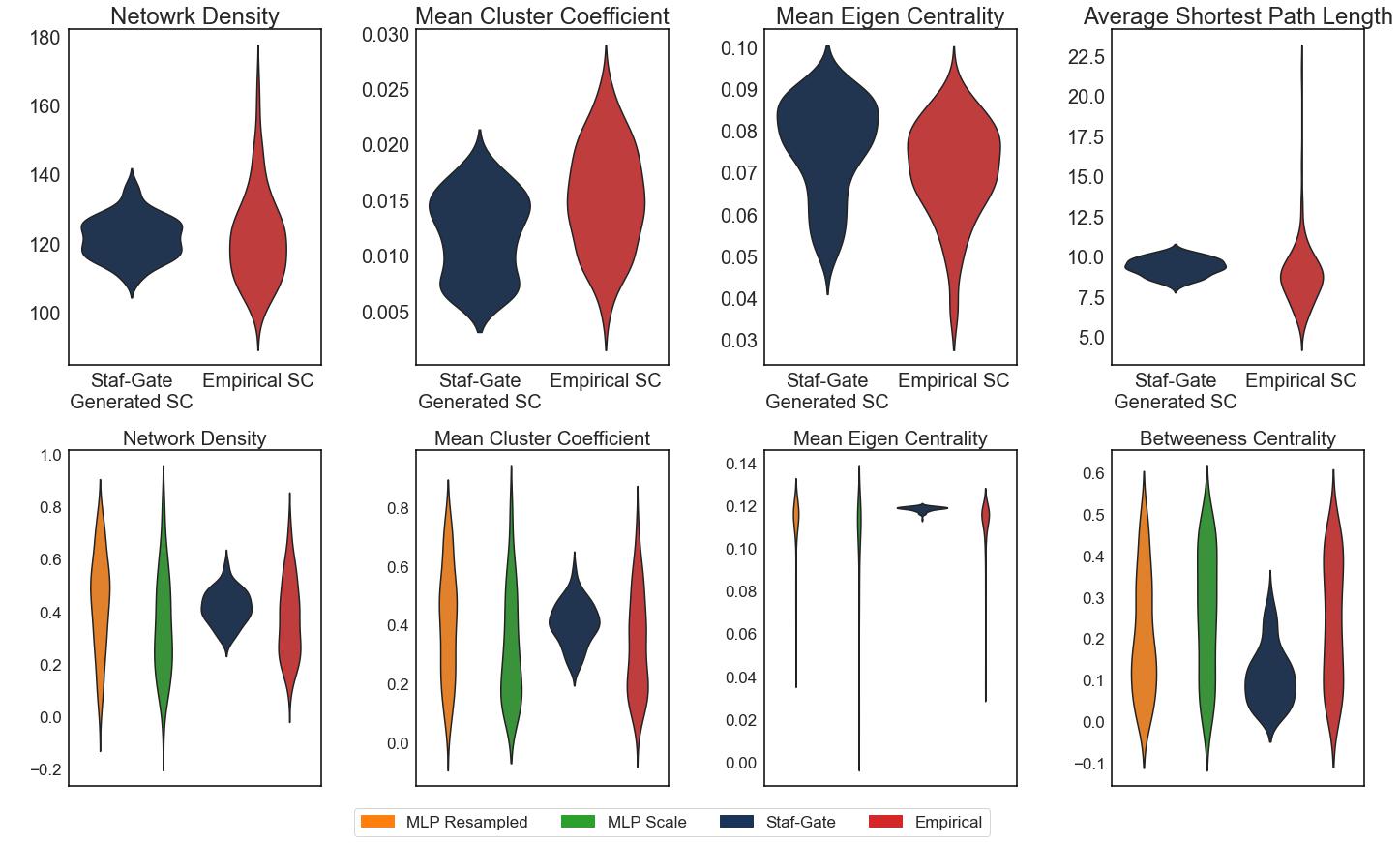}
\label{fig:fc_gen_violin}}
\caption{Goodness of fit analysis of generated SC and FCs. \textbf{(a)} Mean (Left) and Median (Right) of Staf-GATE generated SC (upper-triangular) compare to mean and median of empirical SC (lower-triangular). Mean/median of 100 SCs are generated to compare with the mean/median of 100 empirical SC in the test set. \textbf{(b)} Left: Group Average scatterplot between gFC and eFC, the red dashed line indicates perfect fit. Right: Inter-gFC correlation distribution compared with Inter-eFC correlation distribution. \textbf{(c)} First Row: Generated SC network summary statistics. Second Row: Generated FC network summary statistics. One of the empirical SC is unconnected with infinity path length, which is discarded during plotting.}
\label{fig:generative}
\end{figure}

Empowered by the novel generative architecture, Staf-GATE is capable of generating joint SC and FC pairs by sampling from the approximated posterior $q_{\phi}(z_i|A_i)$ and passing the sampled $z_i$ through the decoders $g$ and $d$; we denote generated SC, FC as gSC, gFC. Similar to previous goodness of fit analysis, we can also study Staf-GATE's efficacy in learning the joint variability of SC and FC through correlation and network statistics. Figure \ref{fig:gen_SC} compares the mean (left) and median (right) eSC to the mean and median of gSC. Group average of gFC, which was predicted by the mean of gSC, was plotted against group average of eFC in Figure \ref{fig:fc_gen_corr} (left) with r=0.95, and inter-gFC correlation is illustrated in Figure \ref{fig:fc_gen_corr} (right).

Similar to the previous results, viewing our generated data as networks allows us to compare the distribution of network summary statistics in Figure \ref{fig:fc_gen_violin}. Staf-GATE achieves good performance in characterizing empirical network summary statistics. Since we modeled SC through scaled data, we rescaled the generated data to the true scale and directly compared the average shortest path length instead of the betweenness centrality. 

We have included binary network analysis for the generated structural connectome in Supplement (refer to Figure \ref{fig:binary} in Supplement \ref{appendix:binary}). To further demonstrate the robustness of our experimental results, we conducted additional model comparisons, which can be found in Supplement \ref{appendix:subcortex} and \ref{appendix:additionalMC}. These results were obtained using a different preprocessing method to acquire SC and FC data. In addition, we incorporated subcortical regions and explored an alternative method of defining SC connection strength (i.e., the fiber count of a connection was normalized by the surface areas of the ROI pairs). The findings are similar to our previous experiments, and therefore, the new results are presented in the Supplement to keep the main paper concise.

\section{Structural Subnetwork Effect on Functional Connectome}
\label{section:subnet}
Our previous experiments suggest SCs and FCs are strongly coupled at both the group and individual levels. Now, we interpret Staf-GATE outputs by finding important SC subnetworks for predicting FCs. We propose a greedy algorithm to interpret our neural network model based on masked inputs, which is inspired by the idea of meaningful perturbations \citep{Fong2017}. In our study, we will perturb the input SC by masking edges (replacing edge weights with 0). We denote a subnetwork as $S$ and use $|S|$ for the number of edges in the upper-triangular adjacency matrix. Masked SCs are denoted by mSC; the corresponding predicted FC is denoted by mFC.

Intuitively, if an SC edge is important for FC prediction, then masking this edge will downgrade the predictive performance. We therefore propose to use mFC-eFC correlation as a loss function and search for subnetworks resulting in large decreases in correlation. We validated the proposed approach by masking hub nodes known to be important for SC-FC coupling \citep{Honey2009, crossley2014} including \textit{left-superior frontal}, \textit{right-insula}, \textit{right-superior frontal}, and \textit{left-precentral}. On average, eliminating these nodes requires masking 238 edges per individual. As a comparison, we created a null distribution for changes in mFC-eFC correlation by masking four random nodes a total of 1000 times. Based on this null distribution, the degradation of masking the hub nodes was found statistically significant ($p \leq 0.05$), indicating that 1) connections to the four hub nodes are important for predicting FCs and 2) the proposed method is an effective method of identifying connections relevant for SC-FC coupling. In practice, masking and comparing all possible subnetworks is infeasible. Instead, we rely on a greedy algorithm that constructs a subnetwork $S$ one edge at a time by iteratively adding the edge that leads to the biggest decrease in predictive performance. The algorithm is terminated once the subnetwork has the desired number of edges. This is formalized in Algorithm \ref{alg:greed}.

\begin{algorithm}
\caption{\newline Interpretation algorithm to search for important subnetworks} \label{alg:greed}
\begin{algorithmic}
\setstretch{1.1}
\State \underline{Input}: eSC, eFC, and subnetwork size $|S_f|$
\State $\text{Initiate $S$ as empty set}$
\State $\text{Initiate mFC}_0 \text{ as pFC; $k$ as 1}$
\While{$k <= |S_f|$}
    \For{each edge $e$ not in $S$}
        \State $\text{Mask all edges in $S$ and the current edge e of eSC}$
        \State $\text{Predict $\text{mFC}_k$ with mSC}$
        \State $\text{Find $e^k_{\max}$ that maximizes } \text{r($\text{mFC}_{k-1}$, eFC)} - \text{r($\text{mFC}_k$, eFC)}$
    \EndFor
    \State $\text{Add } e^k_{\max} \text{ to $S$; Set $k = k+1$}$
\EndWhile \\
\Return{S}
\end{algorithmic}
\end{algorithm}

Applying the greedy algorithm on a $68 \times 68$ symmetric matrix requires $\sum_{i=0}^{|S|} 2278-i$ iterations, which is still computationally burdensome for moderate $|S|$. \cite{Roberts2017CVThres} propose to use the coefficient of variation (defined as standard deviation/mean and denoted as CV) of edge weights to measure the consistency of connections across a population. Edges with a low CV are highly consistent across individuals, whereas those with a large CV are not. Based on their analysis, we applied CV to partition edges into two sets: (1) highly consistent edges ($[0\%,5\%]$ CV among all edges) and (2) moderately consistent edges ($(5\%,15\%]$ CV among all edges). The first set of highly consistent edges was searched for fundamental SC-FC coupling subnetworks in almost all individuals. Subsequently, the set of moderately consistent edges was searched for subnetworks that distinguish SC-FC coupling between groups of individuals with different traits. The robustness of the greedy algorithm is demonstrated in Supplementary \ref{appendix:robust}.

\subsection{Masking Highly Consistent Edges}
\begin{figure}
    \centering
    \includegraphics[width=\textwidth]{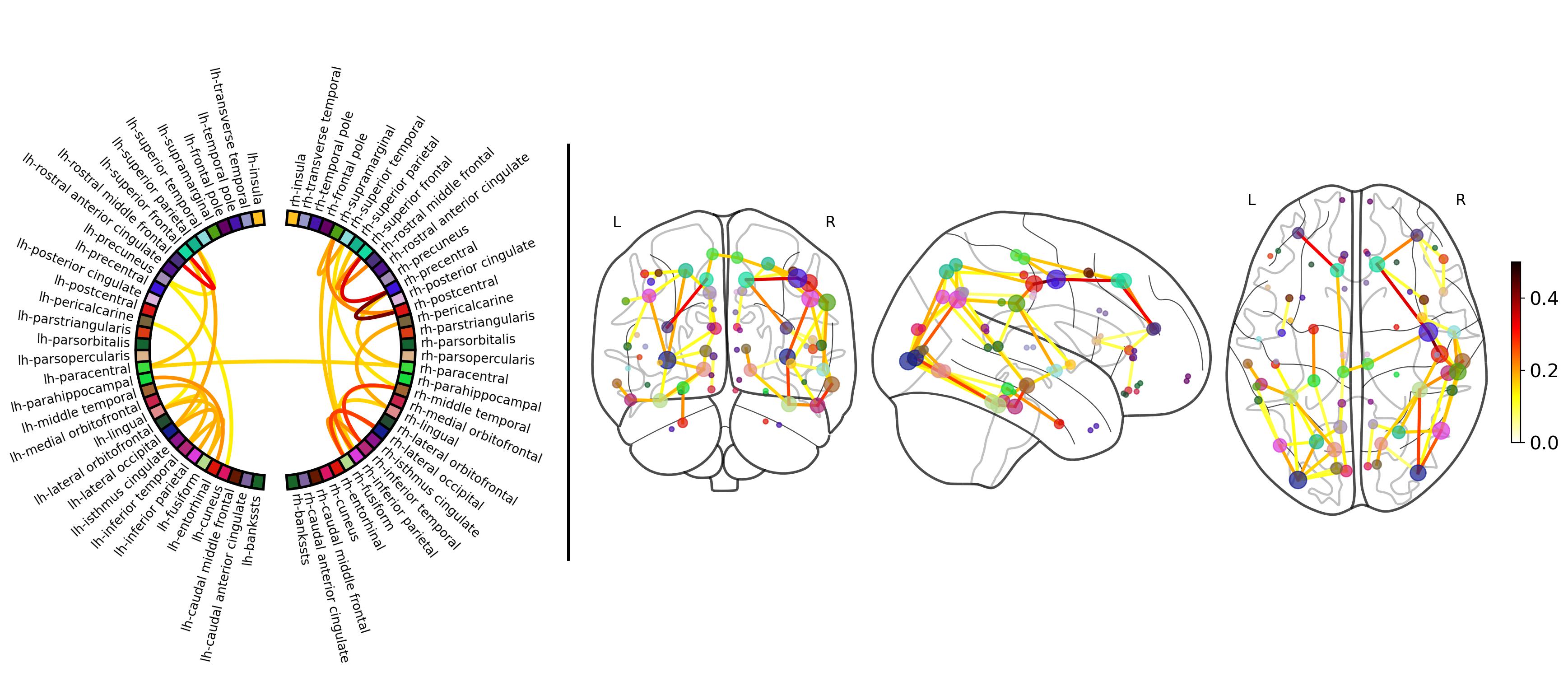}
    \includegraphics[width=\textwidth]{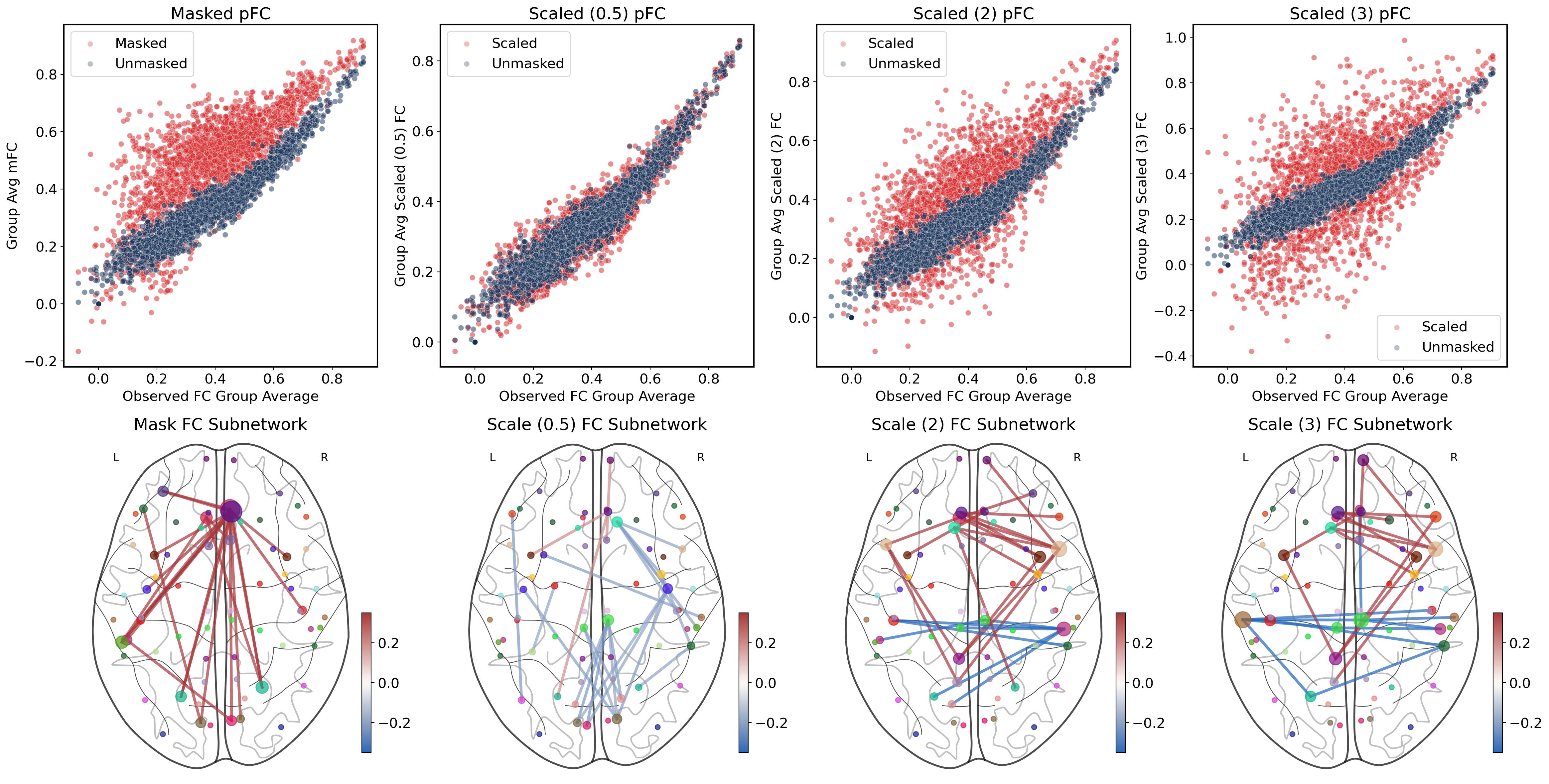}
    \caption{Algorithm determined $S^\star$ and the effects of perturbing $S^{\star}$ on predicted FCs. First Row Left: A partial representation of $S^{\star}$. Top 50 edges of $S^{\star}$ ranked by selection frequency of highly consistent edges. First Row Right: Full subnetwork $S^\star$ mapped on a brain. The color of the nodes in the brain illustration matches the color of ROIs on the circular graph; the size of the nodes in the brain illustration corresponds to the total weighted degree of the ROIs. Second Row: The effect of scaling $S^\star$ on Staf-GATE predicted FC; masking $S^\star$ greatly increases the correlation between regions, but scaling $S^\star$ leads to different deviation. Third Row: Scaling induced deviations from the unmasked prediction mapped to a brain view. }
    \label{fig:common_sub}
\end{figure}

We first study subnetworks containing only highly consistent edges. Applying Algorithm 1 to each individual's SC yields individual-specific coupling subnetworks, which were then aggregated to calculate the selection frequency of each edge. The top 3\% of edges, according to the selection frequency, are collected into a population coupling subnetwork $S^{\star}$ (shown in Figure \ref{fig:common_sub} top left panel). The resulting $S^{\star}$ is important for SC-FC coupling: masking $S^{\star}$ substantially decreased the group average mFC-eFC correlation (from r=0.96 to r=0.78). According to the selection frequency, we can see that \textit{superior frontal} in both hemispheres as well as \textit{right-precentral} and \textit{right-inferior parietal} are important nodes for the SC-FC coupling; stand-alone connections such as \textit{left-paracentral--right-paracentral} and \textit{right-lateral occipital--right-fusiform} are also highly relevant. We relate $S^\star$ to the masking-induced difference in predicted FC. By masking $S^\star$ we observed a major increase of correlation between \textit{right-rostral anterior cingulate}. We further refined our understanding of $S^\star$ by scaling all subjects' edge weights in $S^\star$ by a constant $\delta > 0$: $\delta>1$ boosts connectivity and $0<\delta<1$ reduces it. We observed a highly non-linear relationship between the weights of the subnetwork and predicted FCs, presented in the second and third row of Figure \ref{fig:common_sub} and found that scaling (with either using $\delta>1$ or $0<\delta<1$)   $S^{\star}$ led to a decrease of FC connections of \textit{right-middle temporal} to other ROIs. 

\subsection{Exploration of SC-FC Coupling Difference in Different Groups}
Variations in brain connectivity are known to be important for trait and gender predictions (see e.g., \cite {Durante2017,Liu2021,Ingalhalikar2014, Tunc2016, Tyan2017}), but existing studies mostly focus on one type of brain connectivity. In this section, we explore to use the proposed method to study SC-FC coupling differences in different groups. As an illustration, we study the SC-FC coupling differences in males vs. females.

\begin{figure}
\centering
\includegraphics[width=\textwidth]{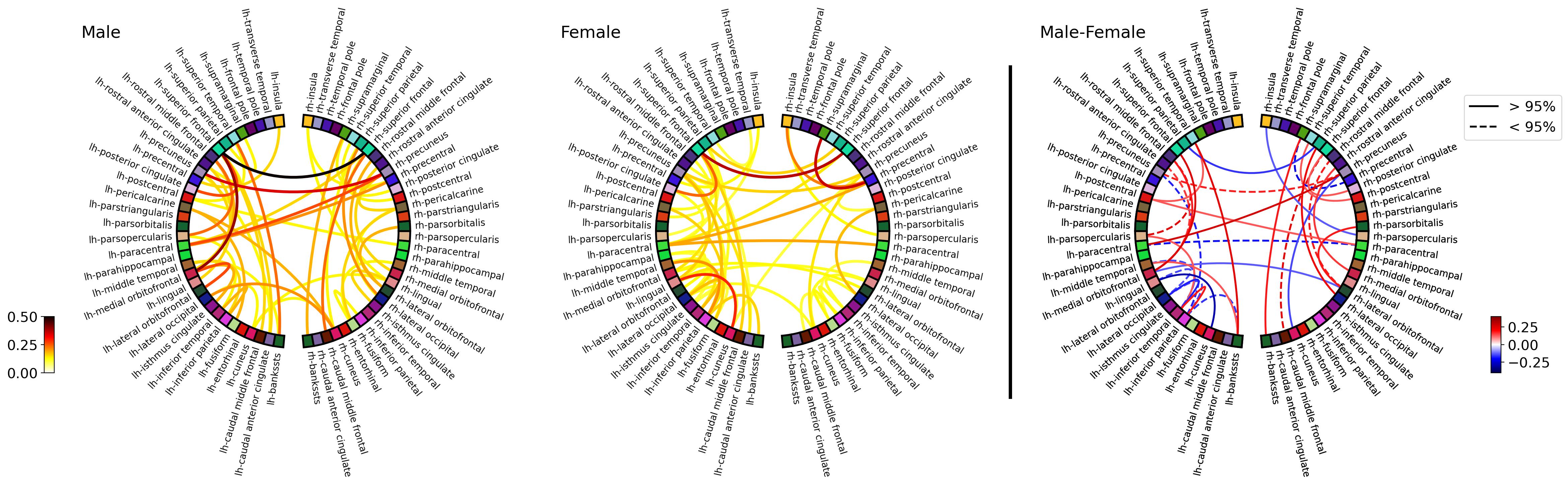}
\caption{Top 50 edges that are the most important for males and females in terms of selection probability and the differences between groups. Left and Center: Top 50 edges in the moderately consistent edge group for males and females' SC-FC coupling. Right: Selection probability difference (top 30 by absolute value) between subgroups ($S_{\text{male}} - S_{\text{female}}$). The solid lines indicate that the connection is statistically significant against the bootstrapped distribution; the dashed line indicates that the connection is not statistically significant. The red (positive) edges indicate that these edges are more relevant for males when compared to females and vice versa for blue (negative) edges. The darker the red (the more positive), the more frequently an edge is selected for the male subgroup compared to the female subgroup and vice versa. }
\label{fig:uncommon}
\end{figure}

We used 32 males and 68 females from the HCP in our study. Our analysis consisted of the following steps to inspect SC-FC coupling differences in male and female groups: 1) we began by identifying the group-specific subnetworks with edges with the top 3\% selection frequency, and denoted the two subnetworks as $S_{\text{male}}$ and $S_{\text{female}}$;  and 2) we took the difference in selection probability $S_{\text{male}} - S_{\text{female}}$. To perform inference, we repeated steps 1) and 2) $30$ times with bootstrapped groups formed with randomly selected subjects to obtain a null distribution for selection probability difference for each edge. We assessed the significance of each edge in $S_{\text{male}} - S_{\text{female}}$ by comparing the group difference in selection frequency to the bootstrap-constructed null distribution. We presented the top 30 edges in $S_{\text{male}} - S_{\text{female}}$ and marked the significant edges (with percentile $>0.95$ according to the null distribution) with solid lines in Figure \ref{fig:uncommon}.

Research has shown that patterns in brain connectivity are associated with sex \citep{Ingalhalikar2014, Tunc2016, Tyan2017,cole2021surface}. We present the sex-group specific networks (i.e., $S_{\text{male}}$ and $S_{\text{female}}$) in the left and center plots of Figure \ref{fig:uncommon}. Comparing the left and center plots, we see that males and females share a large set of SC connections that are important to predicting FC. For example, connections of \textit{left and right superior frontal}, \textit{left-superior frontal and left-posterior cingulate}, and \textit{left-paracentral and right-precentral}. 
The selection frequency difference of $S_{\text{male}} - S_{\text{female}}$ are also presented in the right panel in Figure \ref{fig:uncommon}. We found that 
cross-hemisphere SC connections, such as \textit{left and right-superior parietal} and \textit{left-medial orbitofrontal and right-lateral orbitofrontal}, are crucial for female SC-FC coupling, while connections of \textit{left-paracentral  and right-precuneus} and \textit{left-postcentral and right-paracentral} are important for male SC-FC coupling. 
We also found that within-hemisphere SC connections seem to be more important for male SC-FC coupling. Examples include \textit{left-superior temporal  and left-bankssts}, \textit{right-supermarginal and right-bankssts}, \textit{right-superior frontal and right-caudal medial frontal}, and \textit{right-superior parietal and right-lateral occipital}.
In summary, while our findings on the difference between sex and the corresponding SC can be partially confirmed by previous research on SCs \citep{Ingalhalikar2014, Tyan2017, Tunc2016}, we also discover that cross-hemisphere structural connections are important for males in SC-FC coupling as well.

\section{Discussion}
\label{sec:discuss}
Aiming to leverage the power of neural networks, we generalized graph auto-encoders to study SC and FC coupling. We incorporated brain network topological information and accommodated the skewed FC connection strength distribution in designing our encoders to better characterize the joint distribution of SC and FC. The proposed method achieved state-of-the-art results in predicting both individual FCs (around r=0.6 between eFC and pFC) and the group average FC (with r=0.96). In comparison, for predicting the group average FC using SC, the MLP in \cite{Sarwar2021} achieved r=0.9 while some traditional methods \citep{Honey2009, Messe2014, Misic2016, Rosenthal2018embed} only achieved r=0.7. Attributed to the thoughtful design of the encoders, our model, Staf-GATE, is capable of generating high-fidelity SCs and FCs that accurately mirror the network topology structures of the training data. This opens up wider applications, ranging from the generation of paired (SC, FC) data to SC-FC coupling analysis.

While Staf-GATE demonstrates impressive ability in predicting FC from SC, the intricate relationship between SC and FC remains challenging to comprehend. Recognizing that Staf-GATE is a complicated non-linear model, we employed perturbation-based methods borrowed from the computer vision field \citep{Fong2017} to decode the interplay between SC-FC coupling and other traits. The result masking of highly consistent edges shows that the relationship between SC and FC is highly non-linear. We also studied the SC-FC coupling difference between males and females and showed that important SC connections for predicting FC are different among males and females. This masking procedure is an independent algorithm from Staf-GATE, which can be widely applied to other deep learning models, including MLP methods, to investigate SC-FC coupling outcomes.

Although Staf-GATE has performed well in predicting group-averaged FC from SCs, it falls short in effectively modeling the individual-level relationship. This may indicate that 1) SC only contains limited information about FC, and 2) the current form of Staf-GATE cannot extract sufficiently detailed features from SC to predict the same individual's FC accurately. For 1), other imaging modalities such as EEG, ECG, and MEG may be collected and used to see if they can be used to improve the predictive performance. For 2), more advanced deep learning modules, such as the attention mechanism \citep{Vaswani2017atten}, may be employed to capture better each individual's complicated interactions between SC and FC.

There are also other limitations of Staf-GATE. First, the complexity of  Staf-GATE increases with the dimension of the inputs, making it challenging to study SC-FC coupling using higher-dimensional parcellations. But this limitation is not exclusive to Staf-GATE; it represents a significant drawback of deep learning methods in general, motivating future research directions that treat both SC and FC as continuous functions \citep{cole2021surface}. Second, the generative model in Staf-GATE is based on the VAE framework, which has the drawback of producing high-fidelity images or networks \citep{chen2022inferential}.  To address this, alternative generative algorithms, such as Normalizing Flows, can be considered \citep{Vaswani2017atten}. Third, we still face computational challenges in explaining the SC-FC coupling. For example, we employed the bootstrapping method to generate a null distribution to identify important subnetworks, but such methods are computationally expensive. Note that our interpretation methods exhibit similarities to adversarial machine learning, i.e., they both try to minimize model performance by making minimal alterations to the input. Therefore, techniques employed in machine learning security focused on adversarial scenarios may find application in the interpretation settings of SC-FC coupling.

In future research, we envision several directions to extend the Staf-GATE model. First, Staf-GATE's ability to realistically generate joint SC and FC opens up simulation study and inference opportunities. A pre-trained model for generating SC and FC pairs can be obtained by applying Staf-GATE to large-scale data repositories like HCP. This pre-trained model can then be fine-tuned for small-scale datasets \citep{VanEssen2013,Miller2016,Casey2018}, enabling improved statistical analysis of the small datasets. 
Second, an extension of Staf-GATE could incorporate cognitive traits by introducing an additional layer to generate traits from the latent variable $z_i$. This approach is akin to constructing conditional variational autoencoders, which facilitate interpretable latent structures. 

Third, we can enhance our SC encoder by incorporating graph convolution layers \citep{Kipf16}. 
 Moreover, we will explore methods to enhance our interpretation algorithm, such as employing more efficient optimization techniques to replace the current greedy search. This improvement would yield more stable and less noisy subnetwork outputs.

\nolinenumbers
\section*{Acknowledgement}
This research was partially supported by grant 1R01MH118927-01 of the United States National Institute of Health (NIH). Data collection and sharing for this project was provided by the HCP WU-Minn Consortium (Principal Investigators: David Van Essen and Kamil Ugurbil; 1U54MH091657) which was funded by the 16 NIH Institutes and Centers that support the NIH Blueprint for Neuroscience Research; and by the McDonnell Center for Systems Neuroscience at Washington University. 

\newpage
\bibliography{bib}

\appendix
\renewcommand{\thesubsection}{\Alph{subsection}}
\renewcommand{\thefigure}{\thesubsection.\arabic{figure}}

\renewcommand{\thetable}{A.\arabic{table}}
\renewcommand{\theequation}{A.\arabic{equation}}
\setcounter{equation}{0}
\setcounter{figure}{0}

\setcounter{table}{0}

\newpage
\singlespacing
\linenumbers
\resetlinenumber
\section*{Supplementary Materials}
\subsection{Preprocessing Analysis}
\label{appendix:preprocess}
\begin{figure}
    \centering
    \includegraphics[width=\textwidth]{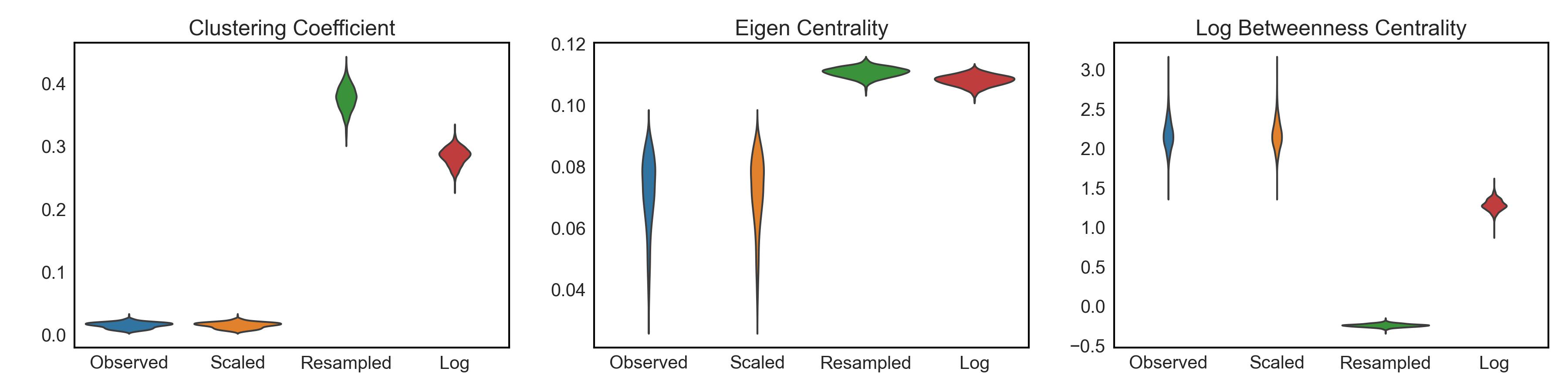}
    \caption{Weighted network summary statistics of the empirical SC, scaled, SC, resampled SC, and log-transformed SC. Gaussian resampling disrupts the distribution of network topological structures while the scaling method preserves these structures. Note that the resulting transformed SCs by different methods are dramatically different in betweenness centrality, leading to near ineligible plot with highly deflated betweenness centrality in Gaussian resampled SC.}
    \label{fig:re_sumstat}
\end{figure}
SC networks generally have an extreme range of edge weights: the median fiber count from the extracted SC is 84; the mean fiber count is 400, but the maximum fiber count is over 30,000. This extreme range of values may cause instability in neural network training \citepsupp{Sarwar2021}. One solution is to apply Gaussian resampling to standardize the data \citepsupp{Honey2009}; another solution is to apply a log-transform to the edge weights. However, Figure \ref{fig:re_sumstat} illustrates that both resampling and the log-transform distort topological characteristics of the brain network. 

To address the unstable training caused by the extreme range of SC entries while preserving the original SC topology, we scaled the SC down by a factor of 100. This preserves the network topology as shown in Figure \ref{fig:re_sumstat}. We tested different scaling factors including unscaled, scaling by 10, and scaling by 100. Using unscaled data or data scaled down by a factor of 10 will lead to overflow during optimization; the scaling factor of 100 gave us stable training resulting in state-of-the-art results.

\subsection{Connections to Other Models}
\label{appendix:connections}

In this section we discuss connections of the proposed Staf-GATE to 1) the latent space model proposed by \citesupp{Hoff2002}, 2) the graph latent factor model by \citesupp{Durante2017}, and 3) the regression Graph Auto-Encoder (reGATE) by \citesupp{Liu2021}.

The latent space model of \citesupp{Hoff2002} 
is designed for modeling a single graph, and assumes conditional independence of the edges given node-specific latent variables.
A related conditional independence assumption is used in Staf-GATE through the latent variable $z_i$. Given the latent variable $z_i$, we assume the elements of $\hat{A}_i$ i.e $A_{i[u,v]}$ and $A_{i[\tilde{u},\tilde{v}]}$, for any $[u,v] \text{ and } [\tilde{u},\tilde{v}], [u,v] \neq [\tilde{u},\tilde{v}]$ are independent\footnote{Similarly for $B_{i[u,v]}$ and $B_{i[\tilde{u},\tilde{v}]}$}. In Staf-GATE, the SC generation follows a Poisson latent factor model, assuming the SC elements are independent Poisson latent variables given $z_i$.

The graph latent factor model proposed by \citesupp{Durante2017} decomposes the adjacency matrix (assuming an undirected graph) into a summation of latent factors similar to Equation \eqref{eq:latent_factor}, where the $X_r^{(i)}(z_i)$ is a latent factor estimated from the latent variable $z_i$. We chose to use this as a generative model for the structural connectomes for the following reasons: 1) the decomposition reduces the number of parameters to estimate as the full SC matrix has $V \times V$ elements while the k latent factor model contains $k \times V$ elements; 2) It is easy to impose topological constraints, for example through the graph KNN layers, to help maintain the topological structure of graphs. The graph latent factor model is only applied to the decoder of SC because FC is a dense matrix that requires much more flexibility to model.

Regression Graph Auto-Encoder (reGATE) is designed for predicting cognitive traits such as reading and vocabulary scores \citepsupp{Liu2021}. Staf-GATE is related to reGATE with both models applying a Poisson latent factor generative model to the decoder, and aiming to learn joint distributions of SC and a response variable. However, Staf-Gate differs from reGATE in three key aspects: (1) the predictive goal: reGATE aims to predict a univariate cognitive trait, whereas Staf-GATE aims to predict a much more complex FC matrix; (2) the predictive network: with different predictive goals, reGATE uses a one layer network, but Staf-GATE uses a deep decoder neural network to realistically characterize high dimensional FC; (3) network invertibility: with a one-layer predictive network, reGATE can infer SCs given cognitive scores by inverting the weight matrix of the predictive layer, but the deep Staf-Gate predictive generator network is non-invertible, which limits Staf-GATE's capability to generate SCs given FCs.

\subsection{Additional Figures for the Simulation Study}

\begin{figure}
    \centering
    \subfigure[Examples of simulated SCs within each group]{
    \includegraphics[width=\textwidth]{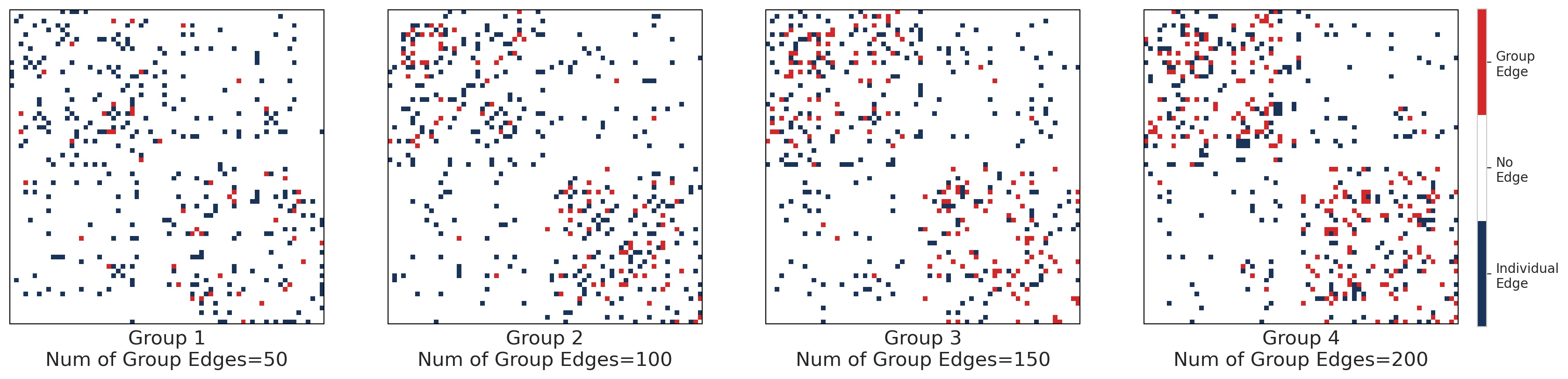}}
    \subfigure[Selected topological network summary statistics for the simulated SCs]{
    \includegraphics[width=\textwidth]{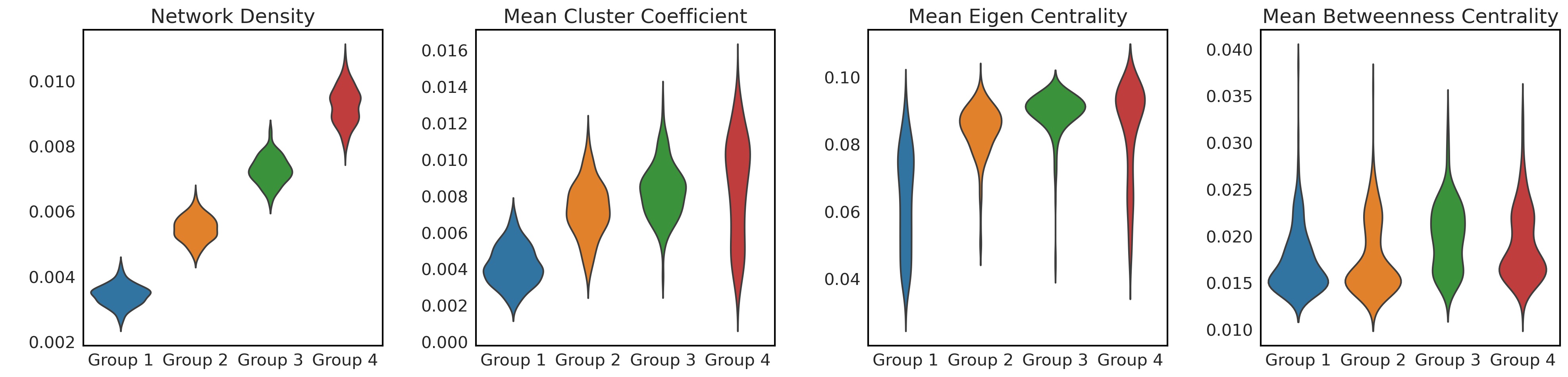}}
    \caption{First row: Examples of simulated SCs in each group. Each heat map corresponds to a simulated SC in a group. The red pixels indicate the group sub-networks $\tilde{A}_y$, and the blue pixels indicate the individual perturbations $E_{yi}$. As the group index increases, the number of group edges, $|\tilde{A}_y|$, increases while the total number of edges stays the same. Second row: Selected network summary statistics of the simulated SC to showcase the different topological structures between each group of $\tilde{A}_{yi}$}
    \label{fig:simu_SC_example}
\end{figure}

In this section, we present additional figures to illustrate the simulated data and results. Figure \ref{fig:simu_SC_example} showcases an example SC from each group and the network topology of our simulated SCs. In particular, as we increase the number of group edges, we observe an increase in weighted density, mean cluster coefficient, and mean eigen centrality, while mean betweenness centrality remains roughly constant. 
\label{appendix:simu}

\begin{figure}
    \centering
    \subfigure[Group average FC for each group]{
    \includegraphics[width=\textwidth]{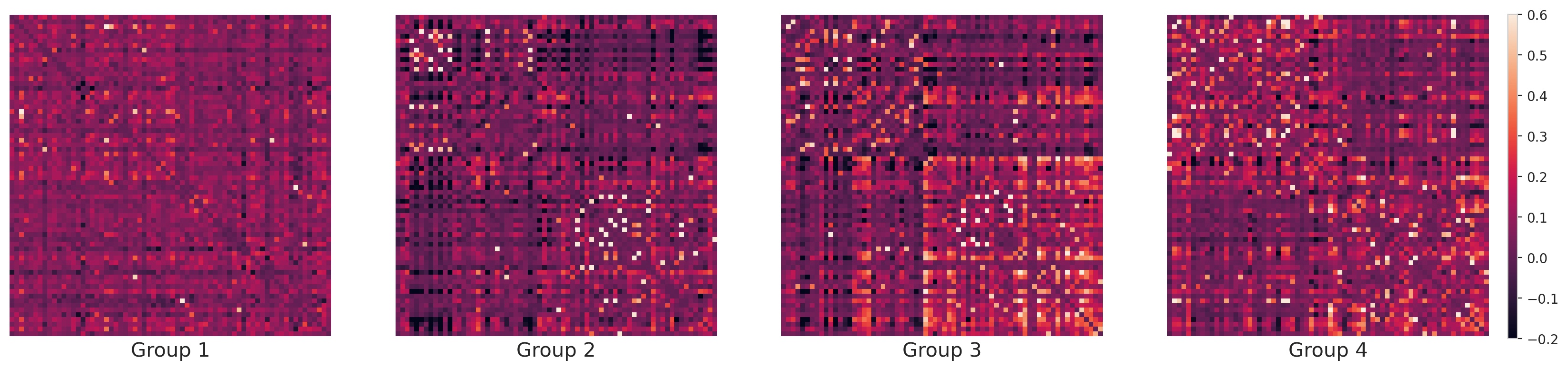}}
    \subfigure[Selected topological summary statistics for the simulated FCs]{
    \includegraphics[width=\textwidth]{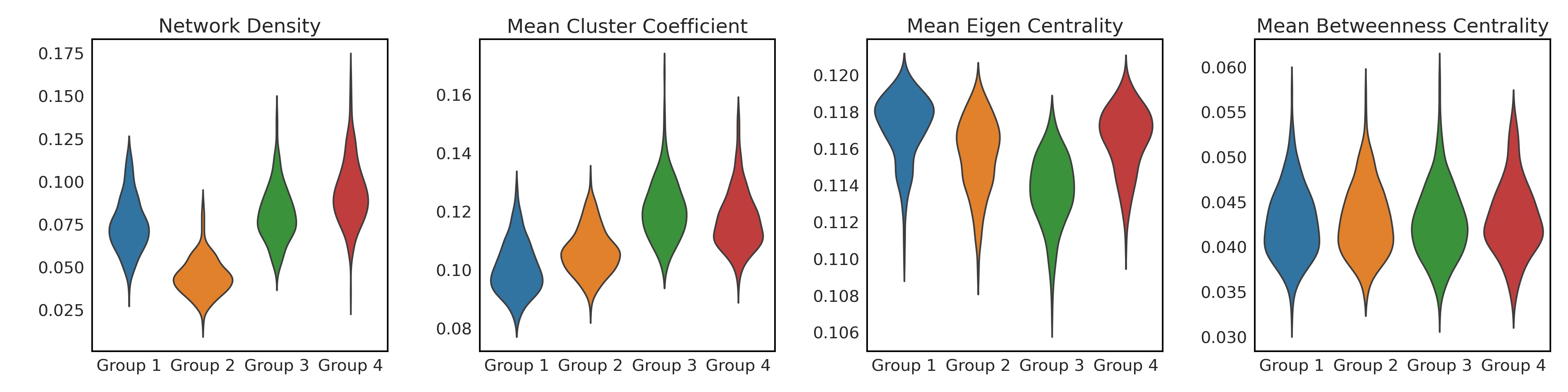}}
    \caption{First row: Each heat map corresponds to the simulated group average $\tilde{B_{yi}}$. The heat maps demonstrate a realistic grid-like structure similar to the empirical FCs. Second row: Selected network summary statistics of the simulated FCs to showcase the different topological structures between each group of $\tilde{B_{yi}}$.}
    \label{fig:simu_FC_example}
\end{figure}

We present the group average simulated FC and the corresponding weighted network topology in Figure \ref{fig:simu_FC_example}. The topology of our simulated FCs has a nonlinear relationship with the topology of the simulated SCs in Figure \ref{fig:simu_FC_example} (b).

\begin{figure}
    \centering
    \includegraphics[width=0.59\textwidth]{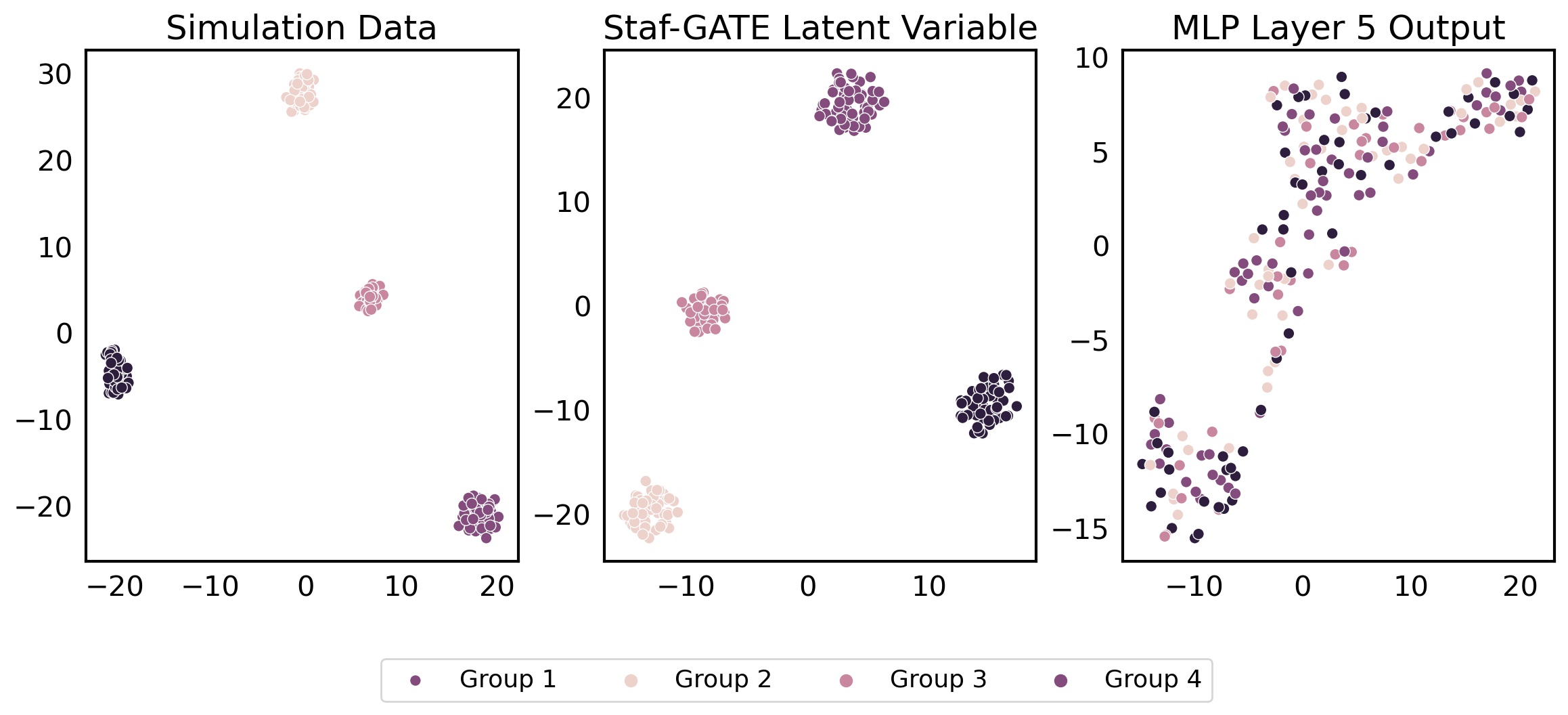}
    \quad
    \includegraphics[width=0.36\textwidth]{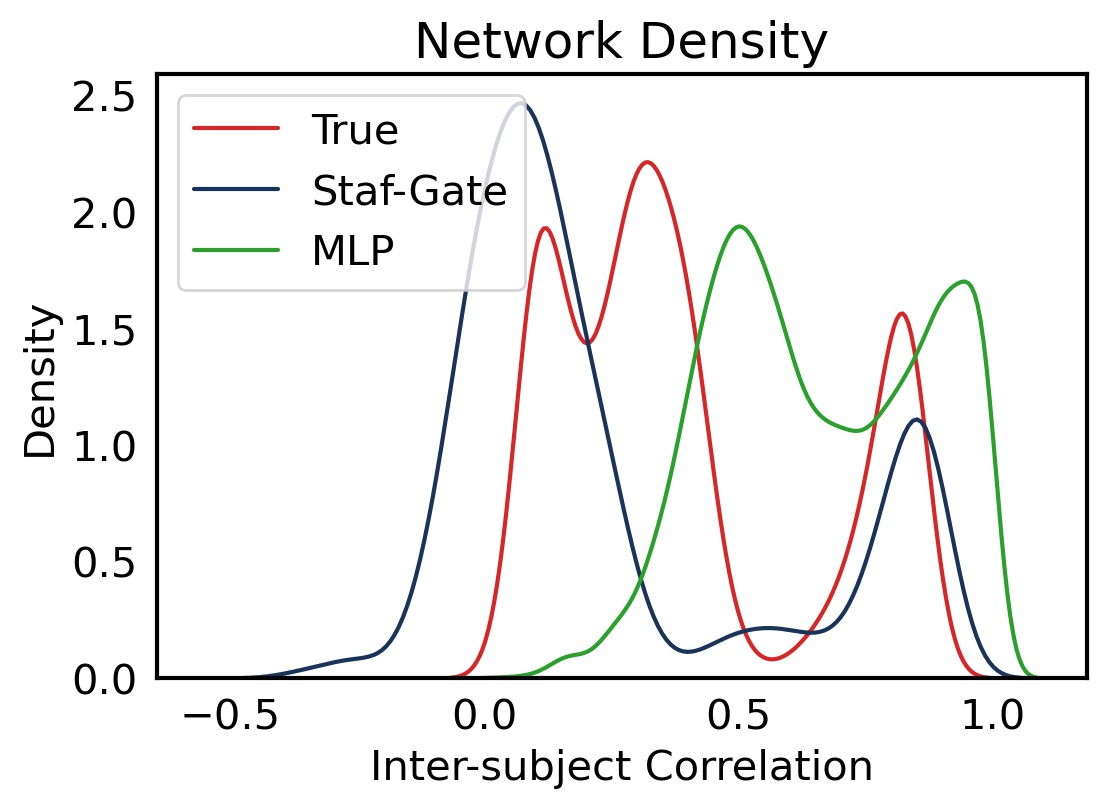}
    \captionof{figure}{Left: Scatter plot of t-SNE reduced input $\tilde{A}_{yi}$, Staf-GATE latent representation $\mu_i$, and the layer 5 output of MLP. Staf-GATE preserves the grouping structure, while MLP does not. Right: Inter-subject correlation of the test samples, Staf-Gate predicted samples, and MLP predicted samples}
    \label{fig:latent_rep}
\end{figure}

Lastly, additional details regarding the latent variable and inter-subject correlation are plotted in Figure \ref{fig:latent_rep}. In Figure \ref{fig:latent_rep} (left panel), we compare the t-SNE reduced SCs, the t-SNE reduced Staf-GATE latent mean, and the layer 5 MLP output. The Staf-GATE latent variable retains the group structure while the MLP latent variable does not. In Figure \ref{fig:latent_rep} (right), we compare the inter-subject correlation between the simulated test and predicted samples. The Staf-GATE predicted samples provide a better representation of inter-subject correlation. 

\subsection{Binary Network Analysis}
\label{appendix:binary}
\begin{figure}
    \centering
    \includegraphics[width=\textwidth]{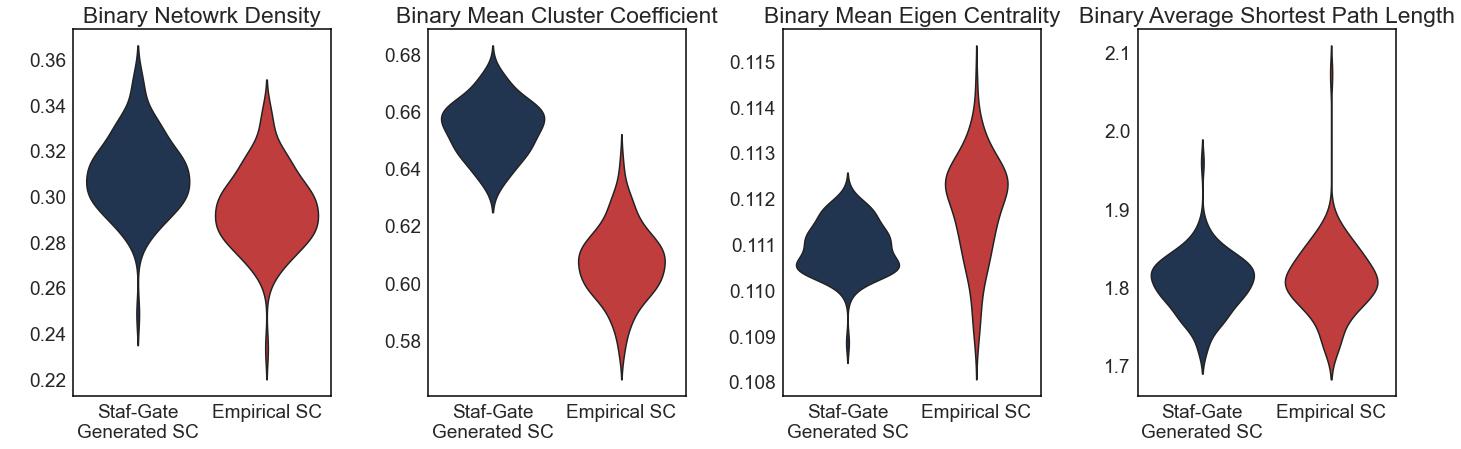}
    \captionof{figure}{Generated structural connectome network topology analysis.}
    \label{fig:binary}
\end{figure}

It is common to reduce weighted adjacency matrices representing SCs to binary adjacency matrices, with a $1$ in entry $i,j$ if regions $i$ and $j$ have any direct connections (regardless of strength) and a $0$ otherwise. We analyzed these binary networks using Staf-GATE; Figure \ref{fig:binary} compares topological summaries of generated networks to empirical networks.

\subsection{More Analysis Including Subcortical Brain Regions}
\label{appendix:subcortex}

\begin{figure}
    \centering
    \includegraphics[width=\textwidth]{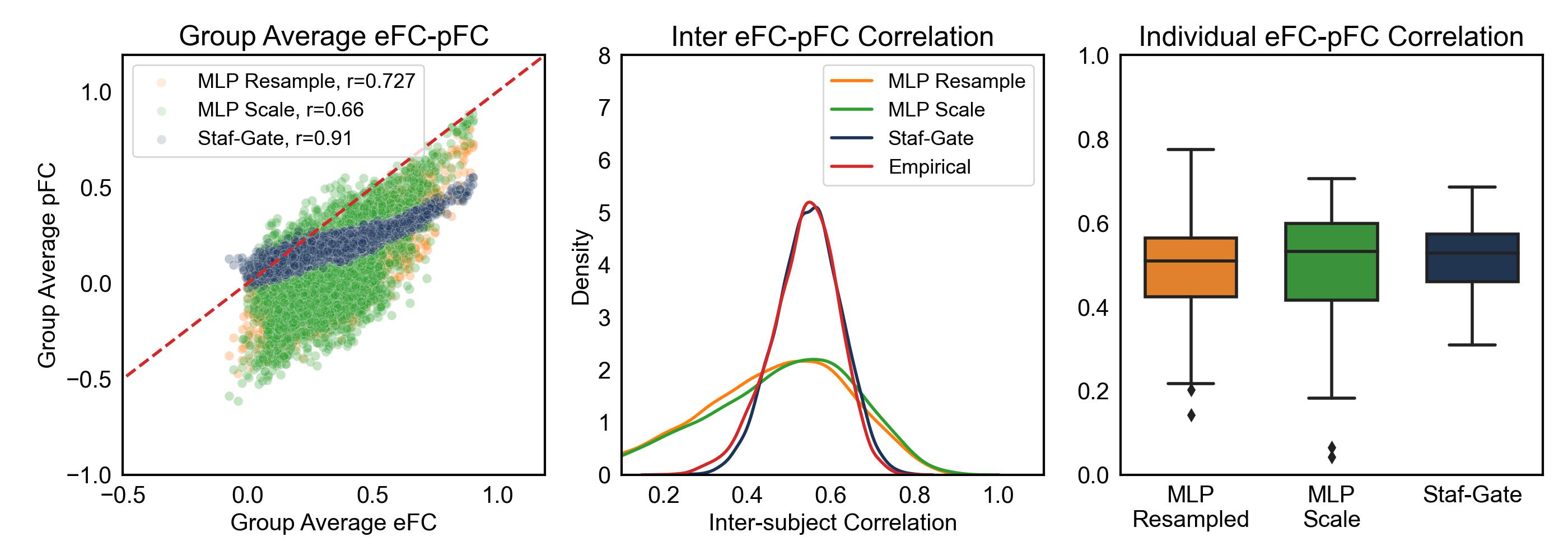}
    \captionof{figure}{Group average eFC-pFC goodness of fit including 19 subcortical regions.}
    \label{fig:subcortex_FC_corr}
\end{figure}

\begin{figure}
    \centering
    \includegraphics[width=\textwidth]{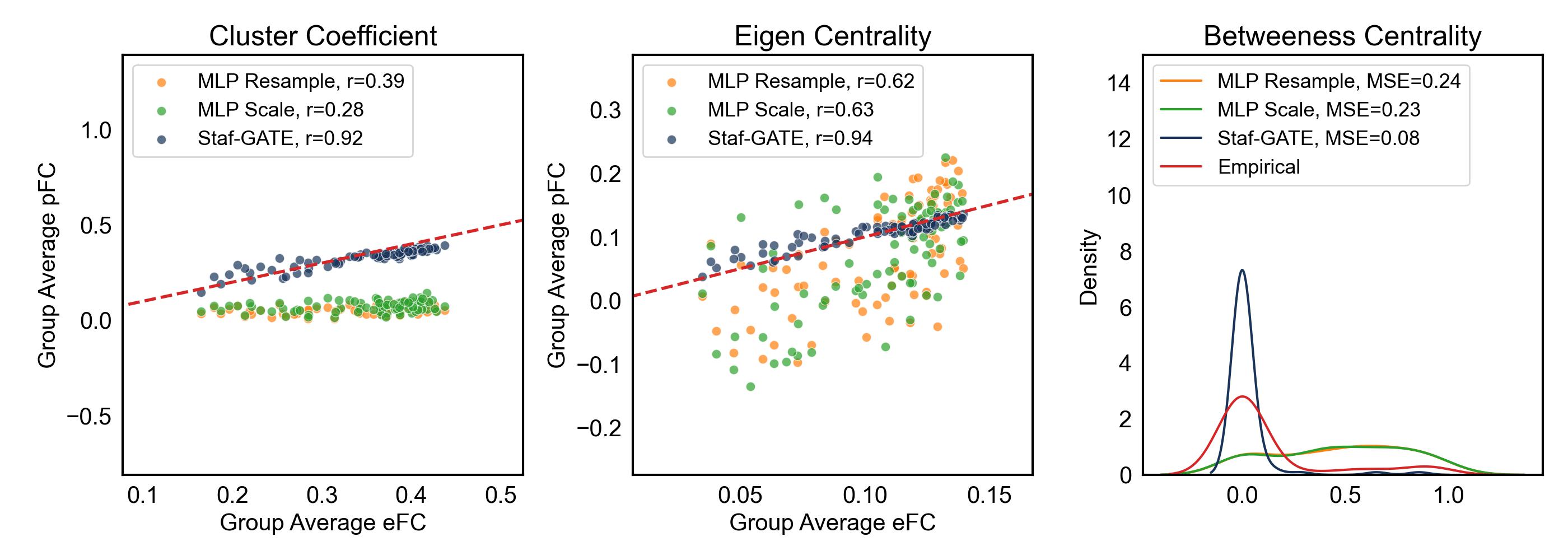}
    \captionof{figure}{Group average eFC-pFC node level topology goodness of fit including 19 subcortical regions.}
    \label{fig:subcortex_top}
\end{figure}

Subcortical regions are highly relevant across many neuroscience applications. We now include 19 subcortical regions (see \citesupp{zhang2018mapping} for more detail of these regions), and compare Staf-GATE to MLP. Our primary focus here is on the topological outcomes and correlation analysis of group averages, as showcased in Figures \ref{fig:subcortex_FC_corr} and \ref{fig:subcortex_top}.

\subsection{Additional Model Comparison Using Data from a Different Preprocessing Pipeline}
\label{appendix:additionalMC}
We conducted additional experiments on data from the HCP using an alternative SC and FC preprocessing pipeline (\citesupp{zhang2018mapping}). A reproducible probabilistic tractography algorithm \citepsupp{girard2014towards, maier2017challenge} was applied to generate the whole-brain tractography data of each subject in HCP. Approximately $10^5$ voxels were identified as the seeding region (the white matter and gray matter interface region) for each individual. About $10^6$ streamlines were generated for each individual, and the Desikan-Killiany atlas was used to derive network nodes. We obtained 1065 subjects with both SC and FC. We then conducted experiments similar to the ones presented in the results section. Additionally, we normalized the streamline count in SC between two ROIs using the surface area to generate a new SC measure and studied whether this normalization step would impact our inference of the relationship between SC and FC. 

\begin{figure}
    \centering
    \includegraphics[width=\textwidth]{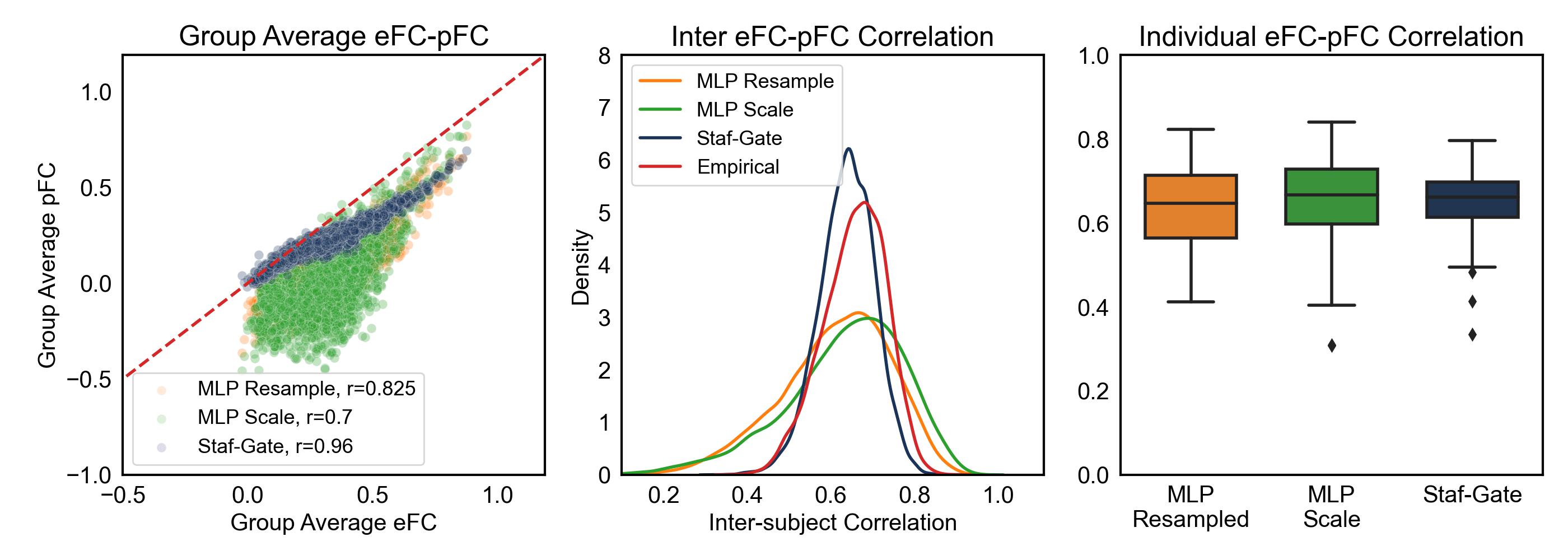}
    \includegraphics[width=\textwidth]{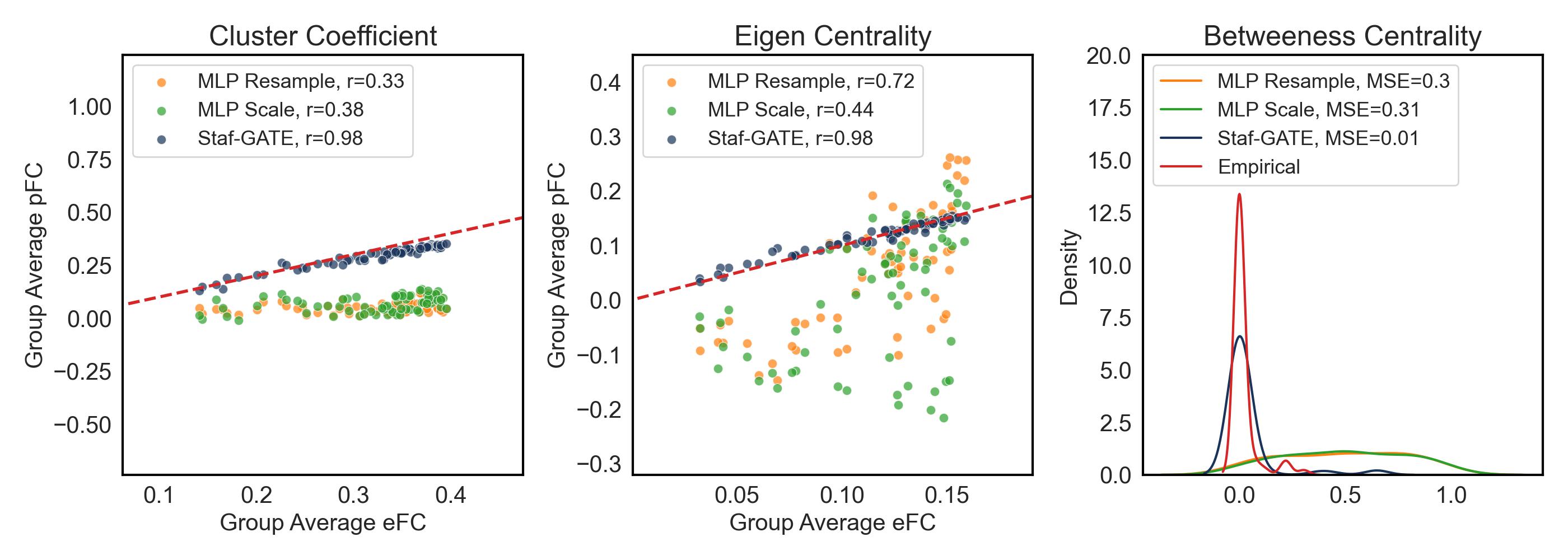}
    \includegraphics[width=\textwidth]{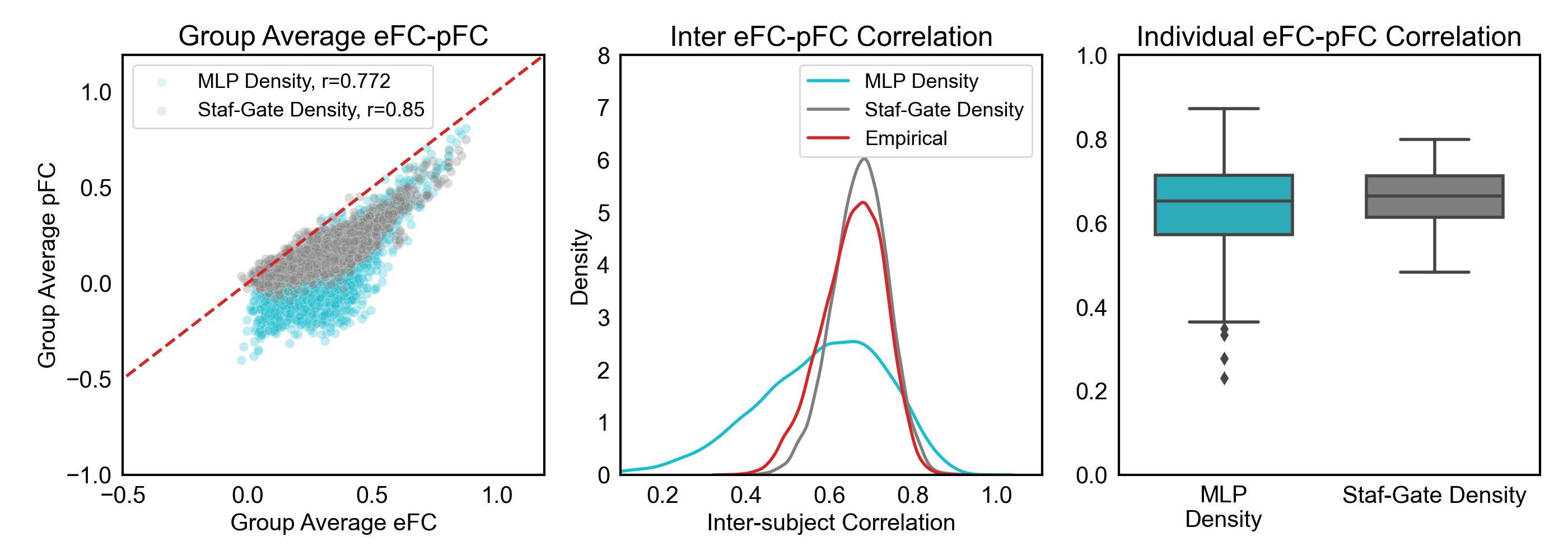}
    \includegraphics[width=\textwidth]{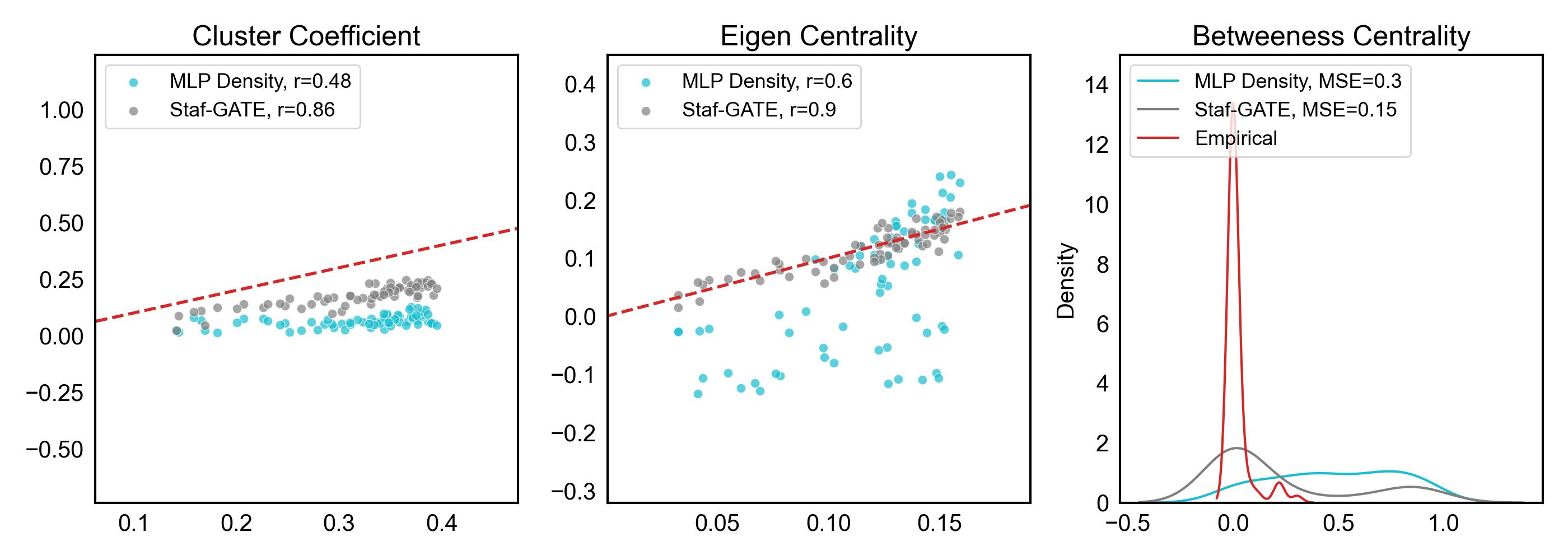}
    \caption{Goodness of fit analysis using data processed through PSC \protect\citepsupp{zhang2018mapping}. First two rows are similar plots as in Figure \ref{fig:FC_corr} with models trained using the PSC pipeline. The third and forth rows are similar plots, but with models trained with the surface-area-normalized SCs from PSC.}
    \label{fig:zhengwu_group_FC_result}
\end{figure}

With these data, we trained MLP using both resampled SC and scaled SC, and we trained Staf-GATE using the scaled SC only. Staf-GATE outperforms both MLP models, as presented in the first two rows of Figure \ref{fig:zhengwu_group_FC_result}. We also included an additional experiment that couples FC using surface-area-normalized SC. The results of the new SC-FC coupling are provided in the last two rows of Figure \ref{fig:zhengwu_group_FC_result}.

\begin{figure}
\centering
\subfigure[Goodness-of-fit assessment of generated SC.]{
\includegraphics[width=0.75\textwidth]{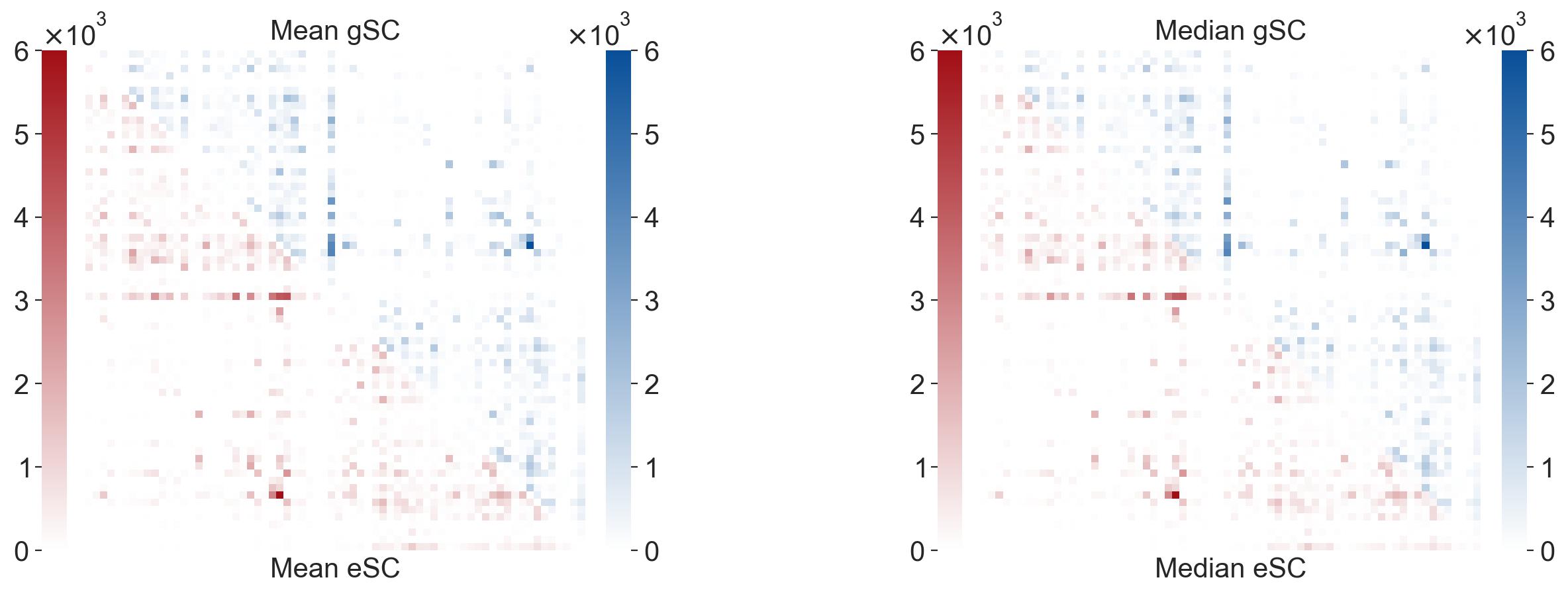}
\label{fig:review_gen_SC}}
\subfigure[Correlation goodness of fit of group average generated FC.]{
\includegraphics[width=0.75\textwidth]{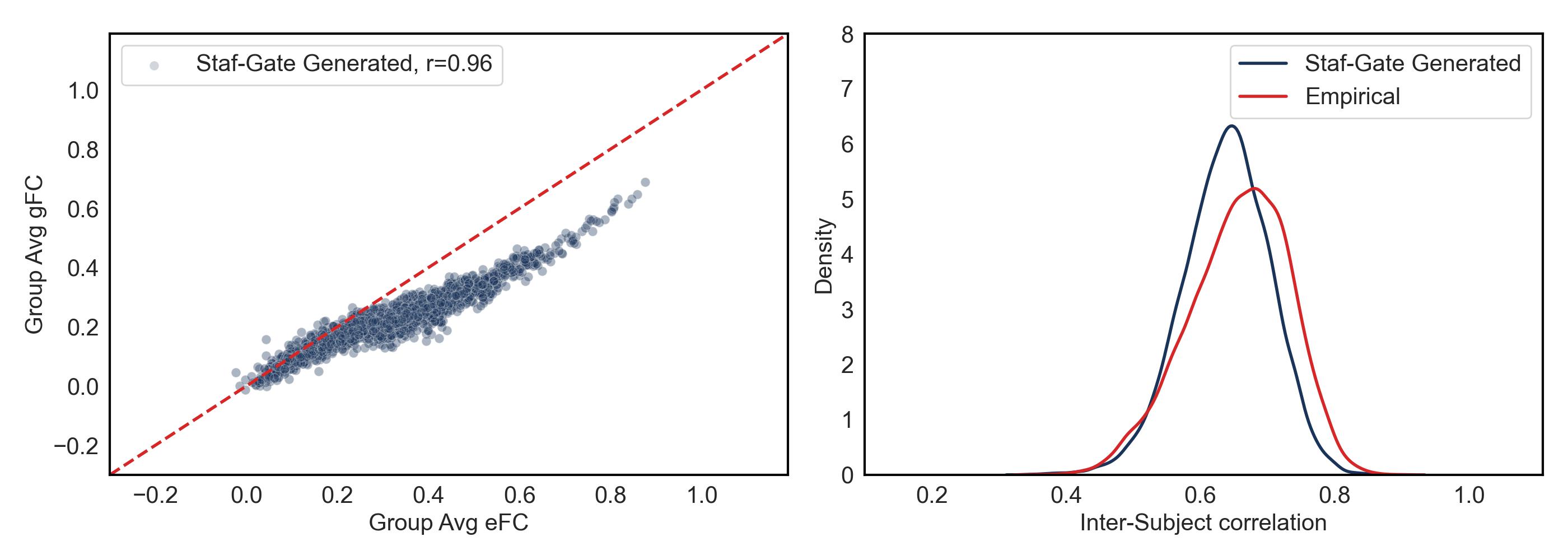}
\label{fig:review_fc_gen_corr}}
\subfigure[Network topology goodness-of-fit assessment for Generated SC and FC.]{
\includegraphics[width=0.75\textwidth]{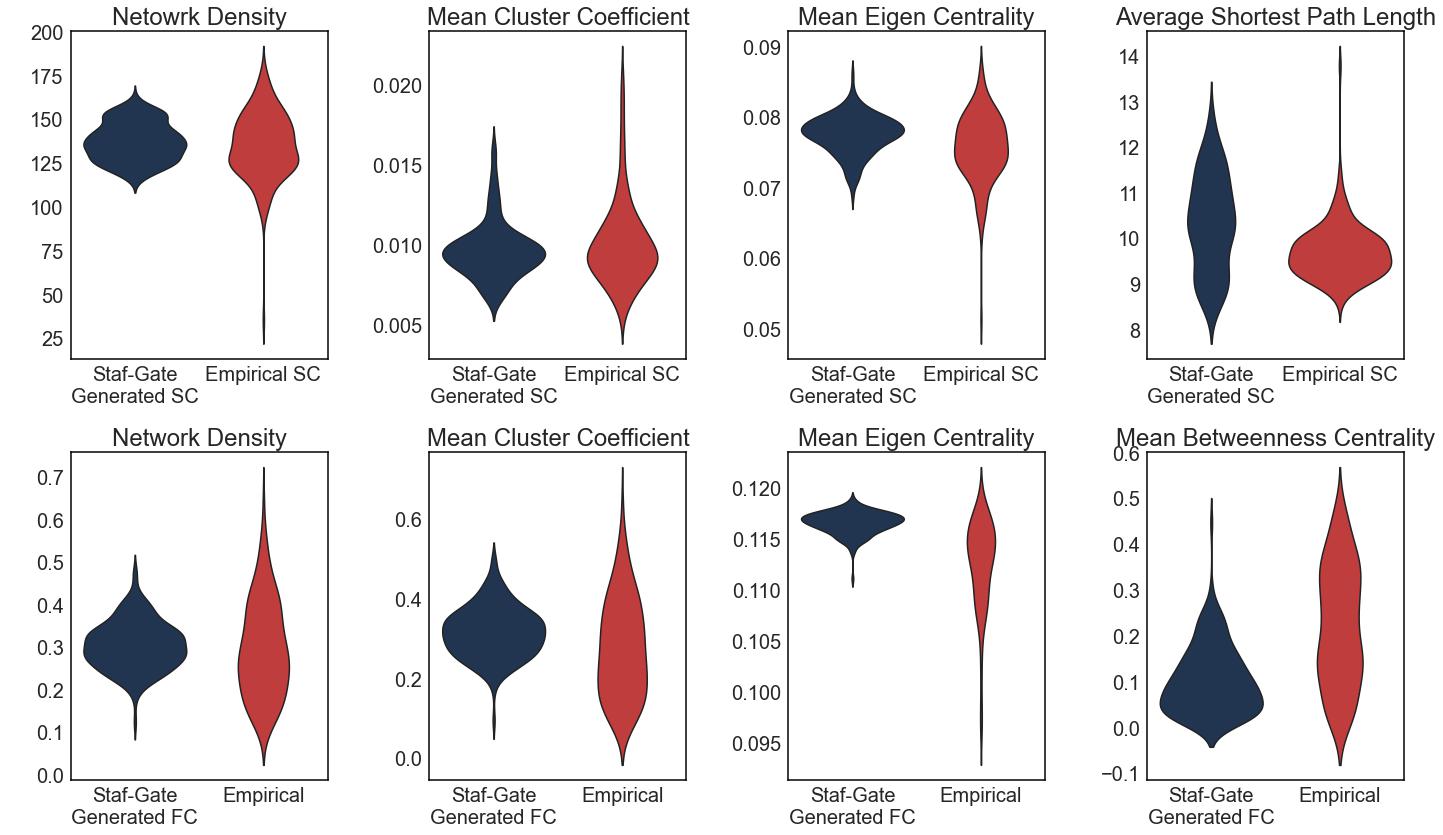}
\label{fig:review_fc_gen_violin}}
\caption{Similar to Figure \ref{fig:generative}, but models were trained  using SCs from the PSC preprocessing pipeline \protect\citepsupp{zhang2018mapping}.}
\label{fig:review_generative}
\end{figure}

\begin{figure}
\centering
\subfigure[Goodness-of-fit assessment of generated SC.]{
\includegraphics[width=0.75\textwidth]{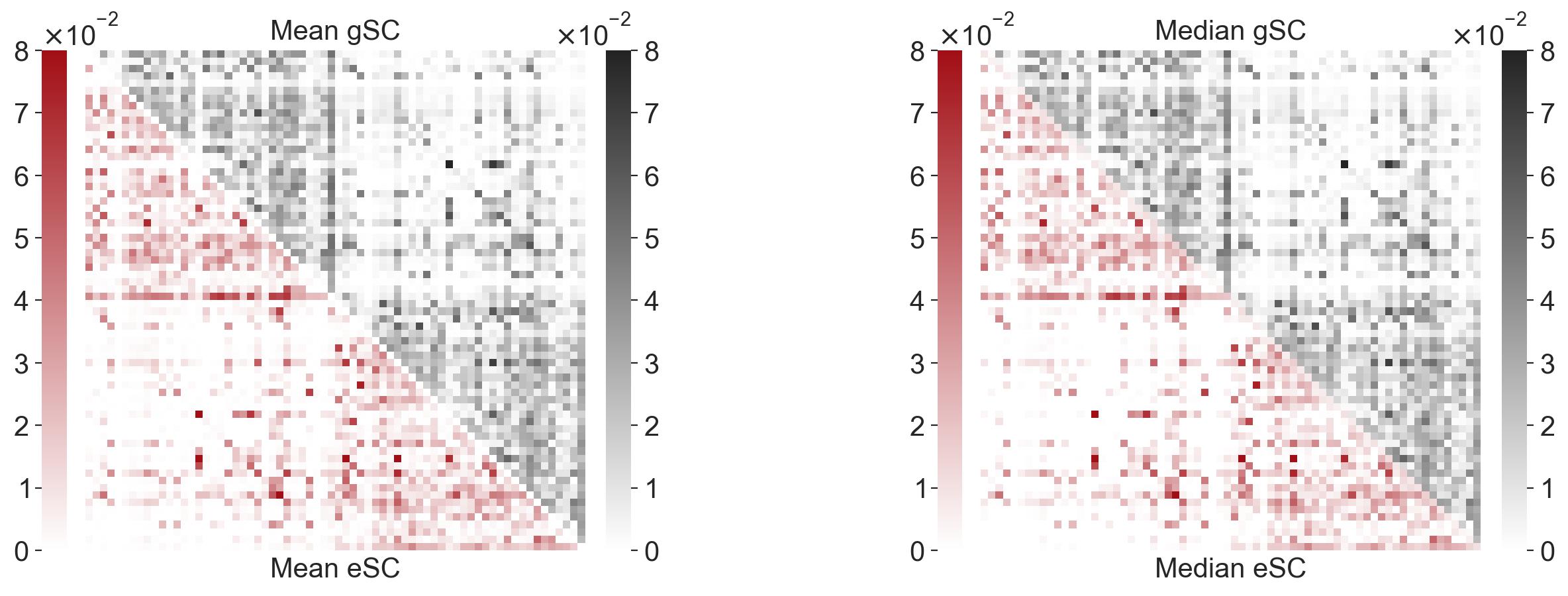}
\label{fig:review_GenSC_EmpSC_density}}
\subfigure[Correlation goodness of fit of group average generated FC.]{
\includegraphics[width=0.75\textwidth]{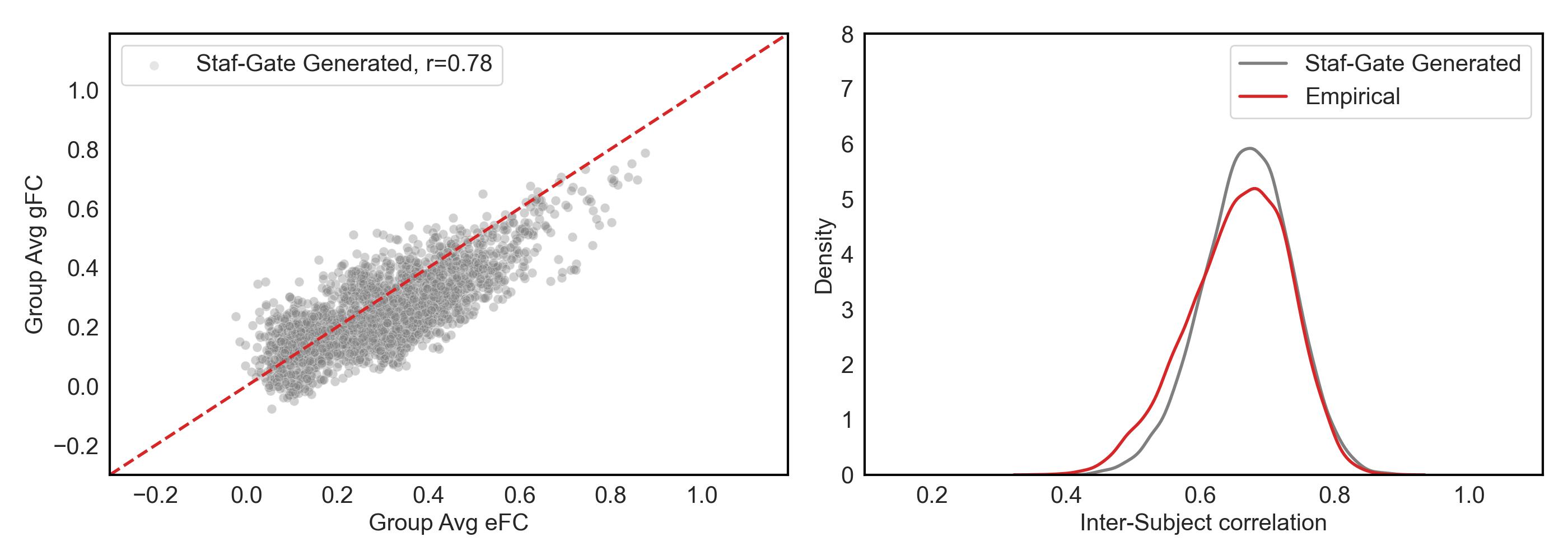}
\label{fig:review_FC_gen_corr_density}}
\subfigure[Network topology goodness-of-fit assessment for Generated SC and FC.]{
\includegraphics[width=0.75\textwidth]{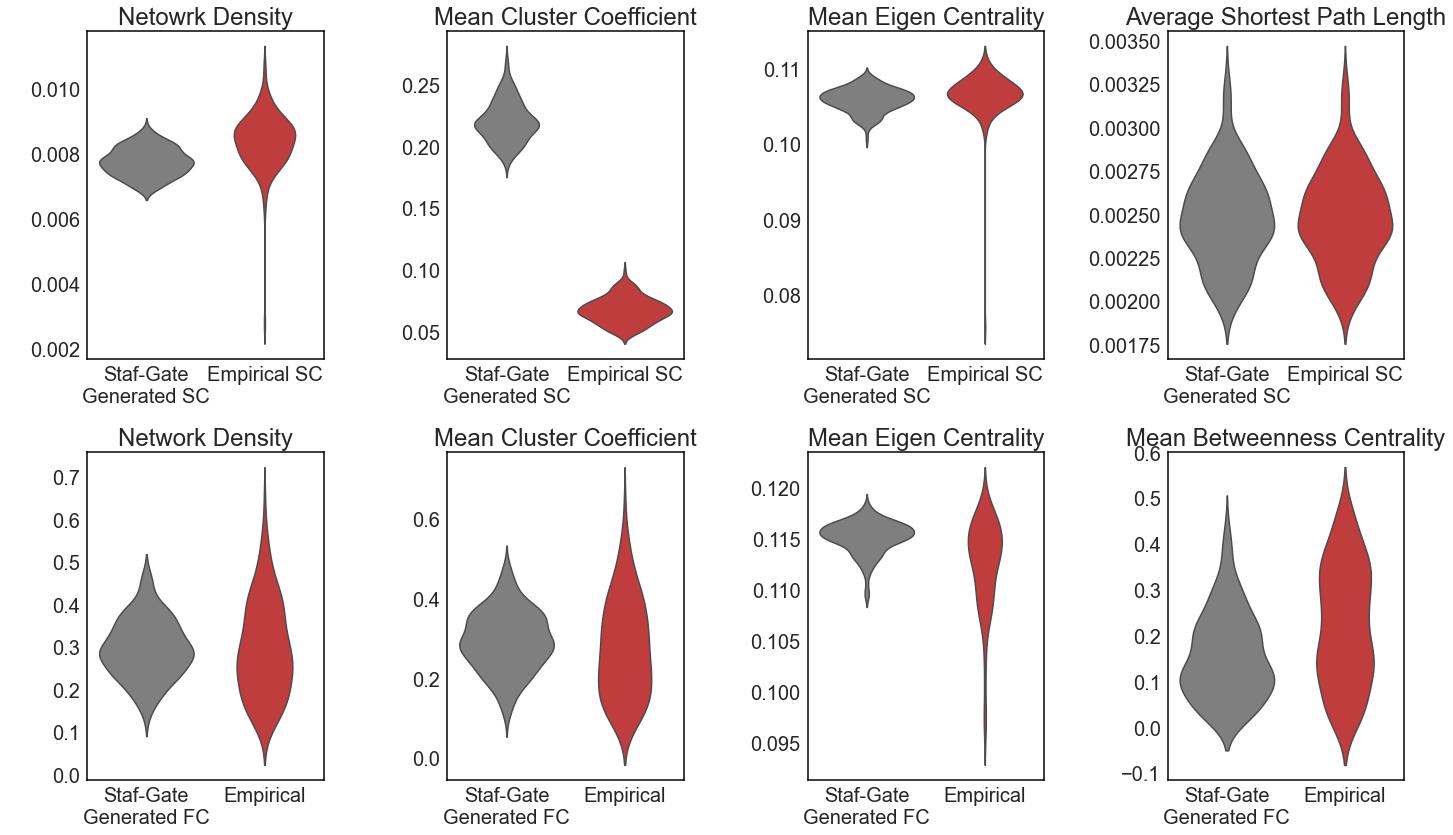}
\label{fig:review_violin_gen_density}}
\caption{Similar to Figure \ref{fig:generative}, but models were trained  using surface-area-normalized SCs from the PSC preprocessing pipeline
\protect\citepsupp{zhang2018mapping}.}
\label{fig:review_generative_density}
\end{figure}

We also analyzed the generative ability of Staf-GATE trained using SCs from the PSC preprocessing pipeline in Figure \ref{fig:review_generative}. Similar results of Staf-GATE trained using surface-area-normalized SCs are presented in Figure \ref{fig:review_generative_density}

\subsection{Tuning and Training result}
\label{appendix:tune}
To tune our model's hyperparameters, we performed a two step grid search. First, we searched in a coarse scale of hyperparameters: learning rate $\in \{1e-4, 2e-4, 3e-4\}$, batch size $\in \{64, 128, 256\}$, $K \in \{40,50,60,70\}$, $\lambda \in \{10,40,70\}$, $c \in \{0.1, 0.3, 0.5, 0.7\}$. From the result of our first grid search, we reduced the range of our search space and perform a second grid search over the reduced space. During the second grid search, we fixed the learning rate at $1e-4$ and batch size at $128$ as this combination showed consistent advantages over the other parameter values. We then fine-tuned the three hyperparameters of our model in a higher resolution grid in the following range: $K \in \{60,70\}$ with an increment of 2, $\lambda \in \{10,40\}$ with an increment of 10, $c \in \{0.3, 0.5\}$ with an increment of 0.05. Through this process, we found the best parameters: learning rate$=1e-4$, batch size$=128$, $K = 68$, $\lambda=20$, $c=0.3$. 

\begin{figure}
    \centering
    \includegraphics[width=\textwidth]{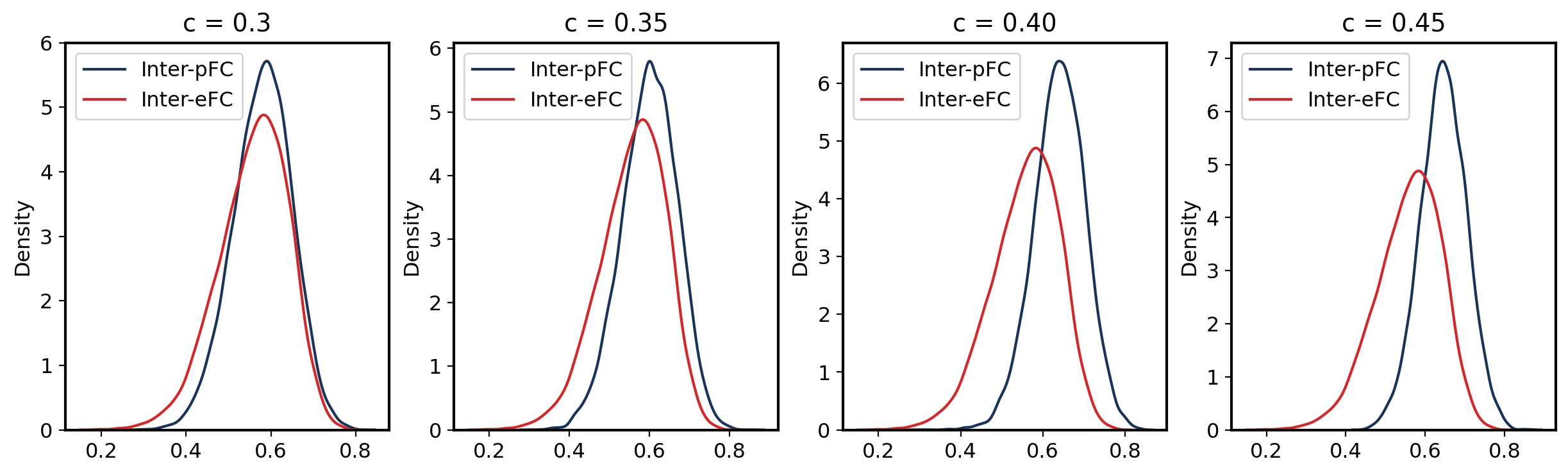}
    \caption{Hyperparameter $c$ selection. As $c$ increases the Staf-GATE predicted inter-subject pFC correlation also increases. Our regularization aims to maintain the inter-subject differences by keeping the inter-subject pFC density as similar to the inter-subject eFC density as possible.}
    \label{fig:tune_c}
\end{figure}

Note that the value of our regularization parameter could affect the loss output during tuning; therefore we followed the heuristic method of regularization parameter selection introduced by \citesupp{Sarwar2021}, especially for $c$. The parameter $c$ controls the inter-subject pFC correlation; we therefore chose this parameter by comparing the validation inter-subject pFC density to the inter-subject eFC density (see Figure \ref{fig:tune_c}).

\subsection{Cognitive Traits Masking Analysis}
\label{appendix:masktrait}
Many different traits, including  demographic, cognitive, and physical measures, were included in HCP data collection \citepsupp{VanEssen2013}. We only used a subset of them in this paper, which includes 1) oral reading recognition score, measuring reading decoding and crystallized abilities, 2) picture vocabulary score, which measures general vocabulary knowledge, 3) line alignment score which measures information processing ability, 
and 4) sex.

We studied SC-FC coupling differences among cognitively high v.s. low groups. We defined a high-scoring group of subjects who are above the median in reading, picture-vocabulary, and line alignment scores simultaneously, and a low-scoring group of subjects who are below the median in all three cognitive scores. This defined 31 high-scoring subjects and 26 low-scoring subjects. We performed inference similar to the steps provided in section \ref{section:subnet}.

\begin{figure}
\centering
\includegraphics[width=\textwidth]{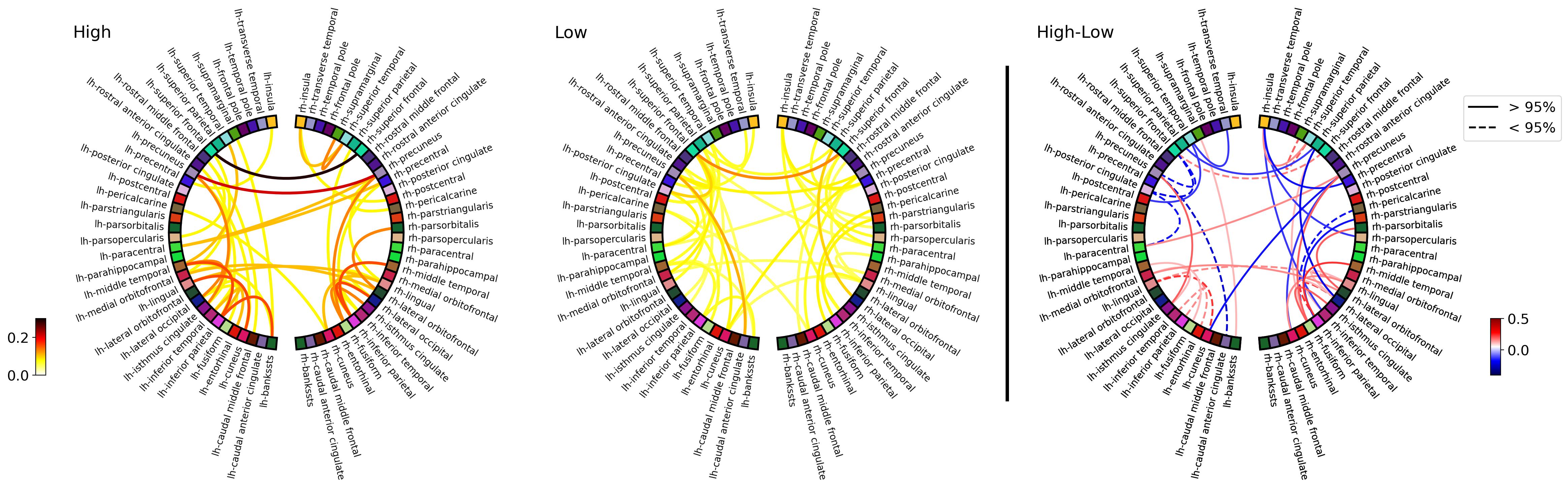}
\caption{Top 50 edges that are the most important for high and low cognitive scoring groups in terms of selection probability and the differences between groups. Left and Center: Top 50 edges in the moderately consistent edge group for high/low cognitive subjects' SC-FC coupling. Right: Selection probability difference (top 30 by absolute value) between subgroups ($S_{\text{high}} - S_{\text{low}}$). The solid lines indicate that the connection is statistically significant against the bootstrapped distribution; the dashed line indicates that the connection is not statistically significant. The red (positive) edges indicate that these edges are more relevant for the high cognitive scoring group when compared to the low scoring group and vice versa for blue (negative) edges. The darker the red (the more positive), the more frequently an edge is selected for the high cognitive scoring subgroup compared to the low cognitive scoring subgroup and vice versa. }
\label{fig:uncommon_trait}
\end{figure}

Figure \ref{fig:uncommon_trait} presents the group-specific top 50 edges and the top 30 edges in $S_{\text{high}} - S_{\text{low}}$. We marked the significant edges (with percentile $>0.95$ according to the null distribution) with solid lines in Figure \ref{fig:uncommon_trait} (right).  
The algorithm tends to select \textit{left-isthmus cingulate}, and \textit{right-fusiform}, \textit{right-precuneus}, and \textit{right-lateral orbitalfrontal} more frequently for the high-scoring group; therefore these nodes are more relevant for high cognition groups' SC-FC coupling. These identified nodes are in line with previous studies on the roles of different regions in cognition \citepsupp{schultz2003role, Cavanna2006precuneus,deen2015functional, Yokosawa2020precuneus}. Low cognition scoring groups, however, have frequently selected edges connected to \textit{left-superior temporal} and \textit{left-superior parietal}. Cross-hemisphere connections including \textit{left-middle temporal - right-lateral orbitalfrontal} and \textit{left-paracentral - right-precuneus} are identified to be important for high-cognitive scoring group's SC-FC coupling. However, some cross-hemisphere connections, including \textit{left-superior parietal - right superior-parietal} and \textit{left-cuneus - right-precuneus}, are important for the low-scoring group. This suggests that there is a limited relationship between the amount of cross-hemisphere connections and different cognitive subgroups' SC-FC coupling.

\subsection{Robustness Analysis}
\label{appendix:robust}
\begin{figure}
    \centering
    \includegraphics[width=\textwidth]{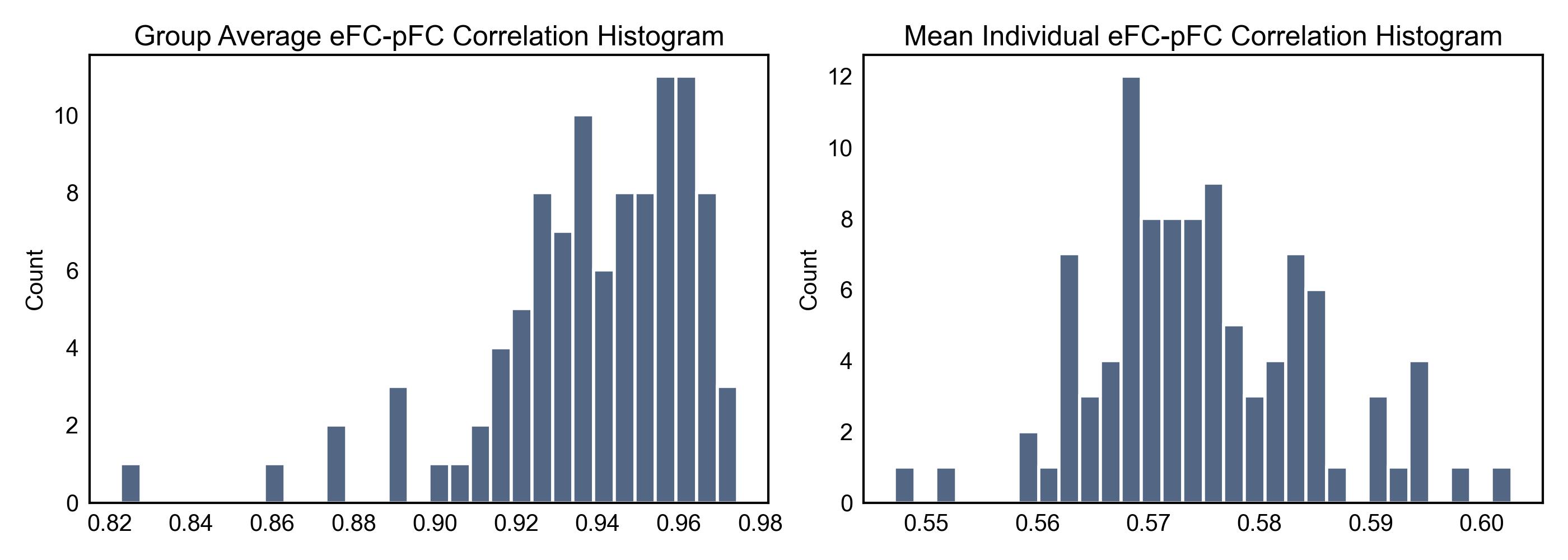}
    \caption{Goodness of fit evaluation of all 100 Staf-GATE model with different initialization and data split. Left: Histogram of group average correlation; Right: Histogram of mean individual correlation.}
    \label{fig:robust_hist}
\end{figure}

We assessed the robustness of our approach by retraining Staf-GATE and running the interpretation algorithm on 100 different data splits, each with a random initialization. We measured predictive performance mainly through correlation between pFC and eFC. In Figure \ref{fig:robust_hist}, we plot the histogram of group average FC correlation (Left) and mean individual FC correlation (Right) for our 100 Staf-GATE models. The predictive performance is consistent with the result presented in the main text.

\begin{figure}
    \centering
    \includegraphics[width=\textwidth]{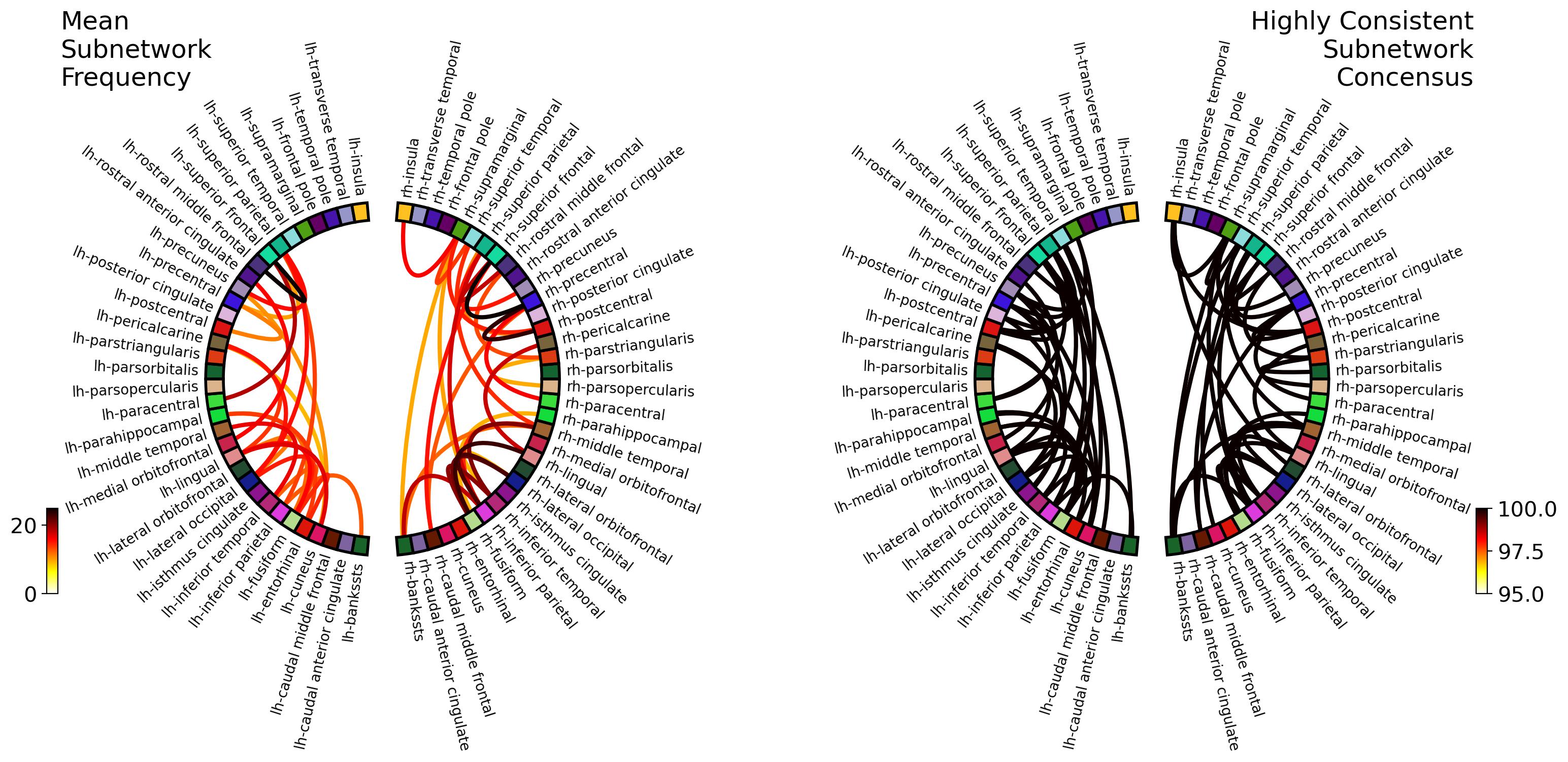}
    \caption{Robustness analysis of our inference applied on highly consistent edges. Left: Mean selection frequency of the edges of Staf-GATE inferred subnetwork. Right: The consensus of 100 Staf-GATE inferred subnetwork on the highly consistent edge set. Our inference is extremely robust as all 100 Staf-GATE has output the same subnetwork as presented.}
    \label{fig:common_robust}
\end{figure}

We also assessed the stability of our interpretation algorithm by applying Algorithm \ref{alg:greed} on 100 different test sets and their corresponding models. Figure \ref{fig:common_robust} presents the mean selection frequency and the consensus network of our 100 different runs. The same SC-FC coupling subnetwork is identified across all runs.
\newpage
\nolinenumbers 
\bibliographystylesupp{model2-names}\biboptions{authoryear}
\bibliographysupp{bib}
\end{document}